\documentclass[fleqn,usenatbib]{mnras}

\usepackage{newtxtext,newtxmath}

\usepackage[T1]{fontenc}
\usepackage{ae,aecompl}
\usepackage{nicefrac}

\usepackage{graphicx}	
\usepackage{adjustbox} 
\usepackage{natbib}

\title[21~cm and metal absorption lines]{Cospatial 21 cm and metal-line absorbers in the epoch of reionization - I : Incidence and observability}
\author[A. Bhagwat et al.]
{Aniket Bhagwat$^{1}$\thanks{E-mail: abhagwat@mpa-garching.mpg.de},
Benedetta Ciardi$^{1}$,
Erik Zackrisson$^{2}$,
Joop Schaye$^{3}$
\\
$^{1}$ Max Planck Institut f\"{u}r Astrophysik, Karl Schwarzschild Stra{\ss}e 1, D-85741 Garching, Germany\\
$^{2}$Observational Astrophysics, Department of Physics and Astronomy, Uppsala University, Uppsala, Sweden\\
$^{3}$Leiden Observatory, Leiden University, Niels Bohrweg 2, NL-2333 CA Leiden, the Netherlands
}

\date{Accepted XXX. Received YYY; in original form ZZZ}

\pubyear{2019}

\begin{document}    
\label{firstpage}
\pagerange{\pageref{firstpage}--\pageref{lastpage}}
\maketitle
%
%
\begin{abstract}
Overdense, metal-rich regions, shielded from stellar radiation might remain neutral throughout reionization and produce metal as well as 21 cm absorption lines. Simultaneous absorption from metals and 21 cm can complement each other as probes of underlying gas properties. We use Aurora, a suite of high resolution radiation-hydrodynamical simulations of galaxy formation, to investigate the occurrence of such "aligned" absorbers. We calculate absorption spectra for 21 cm, {\tt O I, C II, Si II} and {\tt Fe II}. We find velocity windows with absorption from at least one metal and 21 cm, and classify the aligned absorbers into two categories: `aligned and cospatial absorbers' and `aligned but not cospatial absorbers'. While 'aligned and cospatial absorbers' originate from overdense structures and can be used to trace gas properties, 'aligned but not cospatial absorbers' are due to peculiar velocity effects. The incidence of absorbers is redshift dependent, as it is dictated by the interplay between reionization and metal enrichment, and shows a peak at $z \approx 8$ for the aligned and cospatial absorbers. While aligned but not cospatial absorbers disappear towards the end of reionization because of the lack of an ambient 21 cm forest, aligned and cospatial absorbers are associated with overdense pockets of neutral gas which can be found at lower redshift.
 We produce mock observations for a set of sightlines for the next generation of telescopes like the ELT and SKA1-LOW, finding that given a sufficiently bright background quasar, these telescopes will be able to detect both types of absorbers throughout reionization. 
\end{abstract}




\section{Introduction}
The Epoch of reionization (EoR) is the period that sees the cold and neutral intergalactic medium (IGM) being progressively ionized by the first sources of ionizing radiation \citep{gnedin2000,barkana2001,choudhury2005}. Understanding the process of reionization would provide insight into the nature of the first light sources, the evolution of the IGM (see e.g. \citealt{ciardiferrara2005}, \citealt{meiksin2009IGM} and \citealt{Dayal2018} for reviews) and the formation of large-scale structure \citep[e.g.][]{firstgalaxiesloeb13}. Quasar absorption spectra play a key role toward this goal since they probe gas on intergalactic scales \citep[e.g.][]{beckerHighzReview}. The HI Lyman $\alpha$ absorption line is widely used as a tool to study the ionization state of the IGM at low redshifts  \citep[e.g.][]{fan2006,mesinger2010reion,mcgreer2014reion}. At $z > 6$, very limited information can be obtained from Lyman $\alpha$ absorption as the line saturates due to its sensitivity to the neutral hydrogen fraction. However, as discussed by \citet{oh2002} and \citet{furlanetto2003metalabs}, metal absorption lines represent an interesting alternative probe to study the IGM throughout the reionization epoch. 

Metals are produced in stars, and processes such as stellar winds and supernova explosions enrich the IGM around them. Metals in their various ionization states can be observed either as absorption or emission lines. 
As the ionization state of the metal ions is sensitive to various properties of the gas it resides in, like the density, temperature and ionizing radiation field, metal ions can be used to extract a great deal of information about the gas physical conditions as well as the radiation field in the regions of the IGM where they can be observed \citep[e.g.][]{furlanetto2009UVB,finlator2016softUVB,doughty2018alignedUVB,hennawi2021probing}. From the abundance of metals one can constrain the star formation history, whereas column densities, ratios of ionic species and line profiles of absorbers can provide insight into the processes by which galaxies produce and expel metals, the populations of stars that generated them and the ionization state of the metal rich-IGM (see \citealt{beckerHighzReview} for a review). 
    
Absorption lines of metal species such as {\tt O I, C II, Si II, Fe II} \citep[e.g.][]{becker2006OI, mortlock2011quasar, becker2011highZmetalsII, becker2011iron, dodorico2013,bosman2017deep, meyer2019new, becker2019evolution,wu2021oi} and {\tt C IV, Mg II, Si IV} \citep[e.g.][]{WeberRyan2006CIVabs,  simcoe2006CIV, WeberRyan2009downturnCIV, simcoe2011CIVdensity, chen2017mg, cooper2019heavy,dodoricoSiIV22} have been observed in spectra of quasars at the tail end of reionization .
Along with quasar spectra, observations of gamma ray burst (GRB) afterglows also show metal absorption lines in their spectra \citep{chornock2013grb, chornock2014grb, castro2013grb, hartoog2015vlt}. Such observations are complemented by sophisticated numerical simulations, which can provide a very large number of sightlines and thus statistical predictions for absorber properties.
{\tt O I, C II, C IV, Si II, Si IV} are typical lines investigated in the literature (see e.g. \citealt{oh2002,BenOpp2009metallines,garcia2017metalline}). Among these,  {\tt O I} is of particular interest to study during the epoch of reionization as, thanks to its ionization potential of 13.61~eV, it is an excellent tracer of neutral hydrogen (see e.g. \citealt{finlator2013hostOI,keating2013OI,doughty2019evoluOI}).

Another promising probe of reionization is the redshifted 21~cm hyperfine transition of neutral hydrogen (e.g. \citealt{madau199721cm}; see \citealt{furlanetto21cm2006} for a review). Most studies of 21~cm at high redshift focus on the signal in absorption or emission against the CMB \citep[e.g][]{barkana2001, ciardimadau2003, carilli2006hi, furlanetto200621cmglobal, furlanetto21cm2006, pritchard200821cm}. An alternative application of the 21~cm line is as a probe of absorption by the neutral IGM at high redshifts along the line of sight to a distant radio source \citep[e.g][]{carilli2002hi, furlanetto200221minihalo, carilli2004SKA21cm,  carilli2006hi, xu200921cm, xu201121cmgal,ciardi2013,ciardi2015,vsoltinsky2021}. Analogous to the Lyman $\alpha$ forest, the ``21 cm forest" consists of absorption features in the spectrum of a background radio loud source that trace fluctuations in the 21 cm optical depth of the intervening gas. A key advantage of the 21~cm line is that, unlike the Lyman $\alpha$ forest, the 21 cm forest does not saturate at redshifts $z > 6$, allowing it to be, in principle, a much more sensitive probe of small-scale structures (see \citealt{barkana2001, furlanetto21cm2006} for reviews). There has been a great deal of recent progress in observation and instrumentation aimed at the search for the 21 cm signal from the EoR. Ongoing and planned radio observations include the Low Frequency Array (LOFAR)\footnote{\url{www.lofar.org}}, the Experiment to Detect the Global EoR Signature (EDGES)\footnote{\url{http://www.haystack.mit.edu/ast/arrays/Edges}}, the Hydrogen Epoch of Reionization Array (HERA)\footnote{\url{https://reionization.org/}} the Precision Array to Probe the Epoch of Reionization (PAPER)\footnote{\url{http://eor.berkeley.edu/}}, the Murchison Widefield Array (MWA)\footnote{\url{http://www.MWAtelescope.org}} and the Square Kilometre Array (SKA)\footnote{\url{http://www.skatelescope.org/}}. 

Due to the lack of sufficiently bright background sources at the highest redshifts, most studies of metal-absorption at $z>6$ with current 8--10 m telescopes are limited to spectral resolutions $\lambda/\Delta\lambda \lesssim 10000$, which means that weak, narrow metal absorption features are very difficult to detect (but see \citealt{Cooper19} for a couple of such high-resolution detections of metal absorption features against an ultrabright quasar). With upcoming 22-38 m telescopes like the Giant Magellan Telescope\footnote{\url{https://www.gmto.org/}}, the Thirty Meter Telescope\footnote{\url{https://www.tmt.org/}} and the Extremely large Telescope\footnote{\url{https://elt.eso.org/}} (hereafter ELT), high-resolution spectroscopy ($\lambda/\Delta\lambda \approx 100000$) can be applied to point-like sources as faint as $m_{\rm AB} \sim 20$--21 mag, which significantly increases the sample of currently known $z>6$ quasars that can be used for this purpose. In coming years, the Euclid wide-field survey is also expected to produce new detections of quasars in the $z\approx 7$--9 range \citep{Barnett19} that may allow for the discovery of metal absorption systems at higher redshifts than ever before.

At high redshift, we expect to see overdense pockets of gas which are highly neutral due to self-shielding or recombination. Therefore, we can expect metal-enriched but neutral pockets of gas, which, viewed against the light of a background radio loud source such as a quasar or a GRB, could produce a forest of metals as well as 21~cm absorption lines. Thus, 21 cm and metals complement each other as tracers of the neutral IGM, and used in combination (depending on their level of alignment, i.e. their velocity offset) can provide a wealth of information on e.g. underlying gas properties, the location of the densest structures and the gas peculiar velocity. 

In this series of papers we study 21~cm and metal absorption lines throughout the epoch of reionization, using the cosmological radiation-hydrodynamical simulations Aurora \citep{Pawlik_etal_2017}. We aim to investigate aligned metal and 21 cm absorbers to see what information can be extracted from them about the reionization process, early metal enrichment, gas properties and the underlying galaxy populations. In this first paper, we focus on characterizing various scenarios for aligned metal and 21 cm spectra in velocity space, the redshift evolution of their incidence and the observability of these absorbers. We will follow this up with investigations of the galaxy-absorber correlations to study the populations of galaxies traced by the aligned absorbers.

The paper is structured as follows. In Section \ref{sec:Methodology} we discuss our methods -- the simulations used (Section \ref{sec:simulations}), and the calculation of the 21~cm optical depth (Section \ref{sec:21cmtau}). In Section \ref{sec:results} we describe our results -- the physical state of the IGM (Section \ref{sec:PhaseIGM}),  the calculation of synthetic absorption spectra and related results (Section \ref{sec:AbsSpec}) and the classification and incidence of absorbing windows (Section \ref{sec:AlignedAbs}). In Section \ref{sec:Observations}, we discuss mock observations and observability of such a study. Section \ref{sec:conclusions} summarizes our findings. 

\section{Methodology} \label{sec:Methodology}
%
In this section we describe the simulations used to investigate the correlation between HI and metal absorption, as well as the method employed.
%
\subsection{Simulations}
\label{sec:simulations}
%
We use the Aurora suite of radiation-hydrodynamical simulations of galaxy formation during the epoch of reionization \citep{Pawlik_etal_2017}. Here we give some basic information about the simulation and refer the reader to the original paper for more details.

A modified version of the N-body/TreePM Smoothed
Particle Hydrodynamics (SPH) code GADGET \citep{Springel_2005} has been used to perform a suite of cosmological radiation-hydrodynamical simulations of galaxy formation down to redshift $z$ = 6. The hydrodynamics prescription used is the entropy-conserving formulation of SPH \citep{Springel_2005} averaging over 48 SPH neighbour particles. Aurora includes galactic winds driven by star formation and the enrichment of the gas with metals synthesized in stars. The pressure-dependent model of \cite{schaye_vecchia_2008} is used as a sub-grid prescription for star formation. Stellar evolution assumes a \cite{chabrier_2003} initial mass function and determines the element-by-element gas enrichment \citep{wiersma2009}, the rate of supernova events, and stellar ionizing luminosities \citep{schaerer_2003}. Stellar feedback is implemented thermally following the stochastic method of \cite{vecchia_schaye_2012}.  
Aurora tracks the total ejected metal mass\footnote{The metallicity, $Z$, of a particle is defined as the ratio between the mass of elements more massive than helium and the total mass of the particle. Throughout this study we use the solar abundances from \cite{asplund2004solar} to express metallicity values relative to solar ($Z_{\odot} = 0.0122)$.}, and separately, the ejected mass of 11 elements:  H,  He, C, N, O, Ne, Mg, Si and Fe. Ca and S are not tracked explicitly, but rather are assumed to be proportional to Si (see \citealt{wiersma2009}). 

The radiative transfer (RT) of hydrogen and helium ionizing photons is followed using the code TRAPHIC \citep{pawlik_schaye_2008,pawlik_schaye_2011,pawlik_2013,raicevic_2014}, a spatially adaptive ray tracing method which exploits the full dynamic range of the hydrodynamic simulation. This technique allows for capturing the small scale stucture in the ionized fraction within the simulations.

The simulations match a number of observational constraints, including the star formation rate function, as well as the neutral fraction and background photoionization rate at the end of the reionization process.
We specifically use the high resolution L012N0512 simulation which makes use of a box size of 12.5 $h^{-1}$ comoving Mpc with $2 \times 512^3$ dark matter and gas particles. This box reaches a volume-weighted neutral fraction of 0.5 at $z=8.3$. The $\Lambda$CDM cosmological model is adopted with parameters $\Omega_m$= 0.265, $\Omega_b$= 0.0448 and $\Omega_\Lambda$ = 0.735, $n_s$ = 0.963, $\sigma_8$ = 0.801, and \textit{h} = 0.71 \citep{komatsu_2011}.  
\begin{table}
    \centering
    \begin{tabular}{l c c l} 
    \hline
    Ion & $I$ (eV) & $\lambda$ (\AA) & $f$ \\ [0.5ex] 
    \hline\hline
    O I    & 13.61     & 1302.16    & 0.048  \\ 
    C II   & 24.39     & 1334.53    & 0.127  \\ 
    Si II  & 16.34     & 1260.42    & 1.180  \\ 
    Fe II  & 16.18     & 2382.76    & 0.320  \\  [1ex] 
    \hline
    \end{tabular}
\caption{Metal absorption lines investigated in this study. From left to right the columns refer to: the ion, its ionization potential, $I$, rest frame wavelength, $\lambda_0$, and oscillator strength, $f$, of the transition.}
\label{table:transitions} 
\end{table}
In this study, we will focus on absorption from oxygen, carbon, silicon and iron, as they have strong resonance lines redward of the Ly$\alpha$ line, which make them viable to be observed at high redshifts. We refer the reader to Table~\ref{table:transitions} for the lines and their characteristics.
Since the ionization potential of the {\tt OI} line is only 0.02 eV higher than that of {\tt HI}, the two can exist in tight charge exchange equilibrium \citep{Osterbrock}, and therefore, as suggested in previous studies \citep{oh2002,Oxypump2007,keating2013OI,doughty2019evolution}, the {\tt O I} line can act as a tracer for neutral hydrogen.
Lines of other metals such as {\tt C II}, {\tt Si II} and {\tt Fe II} are also promising absorption lines that are well known to be present in neutral gas (e.g. \citealt{becker2006OI,becker2011highZmetalsII}).
\subsection{21~cm optical depth} 
\label{sec:21cmtau}
The 21~cm line corresponds to the hyperfine, spin-flip transition of the ground state of the neutral hydrogen atom, at rest frame frequency $\nu_{\rm 21cm}$ = 1.42 GHz.  The 21~cm optical depth of a patch of neutral hydrogen located at redshift $z$ along the line of sight can be written as \citep{semelin_2016}:
\begin{equation}
\begin{split}
    \tau_{\rm 21cm} = \frac{3}{32\pi}\frac{{h_p}c^3A_{10}}{k_B {\nu_{\rm 21cm}}^2}\frac{n_{\rm HI}}{H(z) T_S}\left(1+\frac{1}{H(z)}\frac{dv_{||}}{dl}\right)^{-1} \\
    \approx 0.0088 (1+z)^\frac{3}{2} \frac{\delta \ x_{\rm HI}}{T_S}\left(1+\frac{1}{H(z)}\frac{dv_{||}}{dl}\right)^{-1}
\end{split}
    \label{Eq:21cmtau}
\end{equation}
where $A_{10} = 2.85 \times 10^{-15} s^{-1}$ is the Einstein coefficient of the corresponding hyperfine transition, $h_p$ is the Planck constant, $H(z)$ is the Hubble parameter, $n_{\rm HI}$ is the neutral hydrogen number density, $T_S$ is the hydrogen spin temperature, $\delta$ is the baryon overdensity\footnote{We define overdensity $\delta = \rho/\overline{\rho}$, where ${\rho}$ is the gas density and $\overline{\rho}$ is the mean baryon density at redshift $z$.} and $x_{\rm HI}$ is the neutral hydrogen fraction. Finally, $\frac{dv_{\parallel}}{dl}$ is the peculiar velocity gradient along the line of sight, with $dv_{\parallel}$ and $dl$ both in proper or comoving units. The other symbols appearing in the equation have the standard meaning adopted in the literature. We make the common assumptions that $(1 + \frac{1}{H(z)} \frac{dv_\parallel}{dl})$ $\approx1$
and that the spin temperature is coupled to the kinetic temperature of the gas.
\section{Results}
\label{sec:results}
\subsection{Physical state of gas}
\label{sec:PhaseIGM}
%
\begin{figure}
    \centering
	\includegraphics[width=\columnwidth]{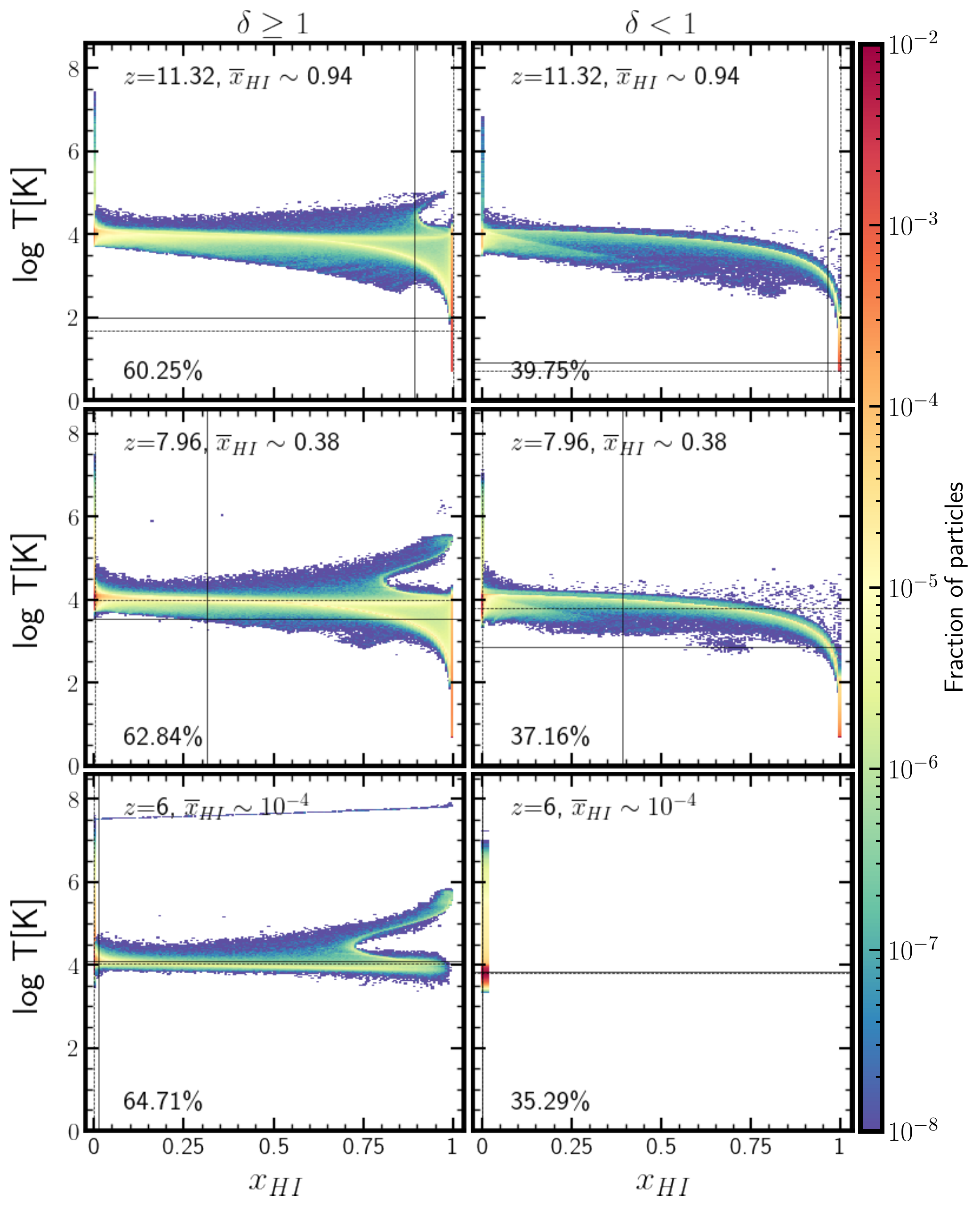}
	\caption{Distribution of gas temperature as a function of neutral hydrogen fraction $x_{\rm HI}$ at $z$ = 11.32 (top panels), 7.96 (middle), 6.0 (bottom) for overdense (left column) and underdense (right) particles. $\overline{x}_{\rm HI}$ indicates the volume-weighted neutral fraction at each redshift. The horizontal solid (dashed) lines indicate the mean (median) temperatures of the particles, while the vertical solid (dashed) lines indicate the mean (median) neutral hydrogen fraction. Percentages in each panel indicate the fraction of particles in an overdense/underdense state.}
	\label{fig:Txhhist}
\end{figure}
\begin{figure}
    \centering 
	\includegraphics[width=\columnwidth]{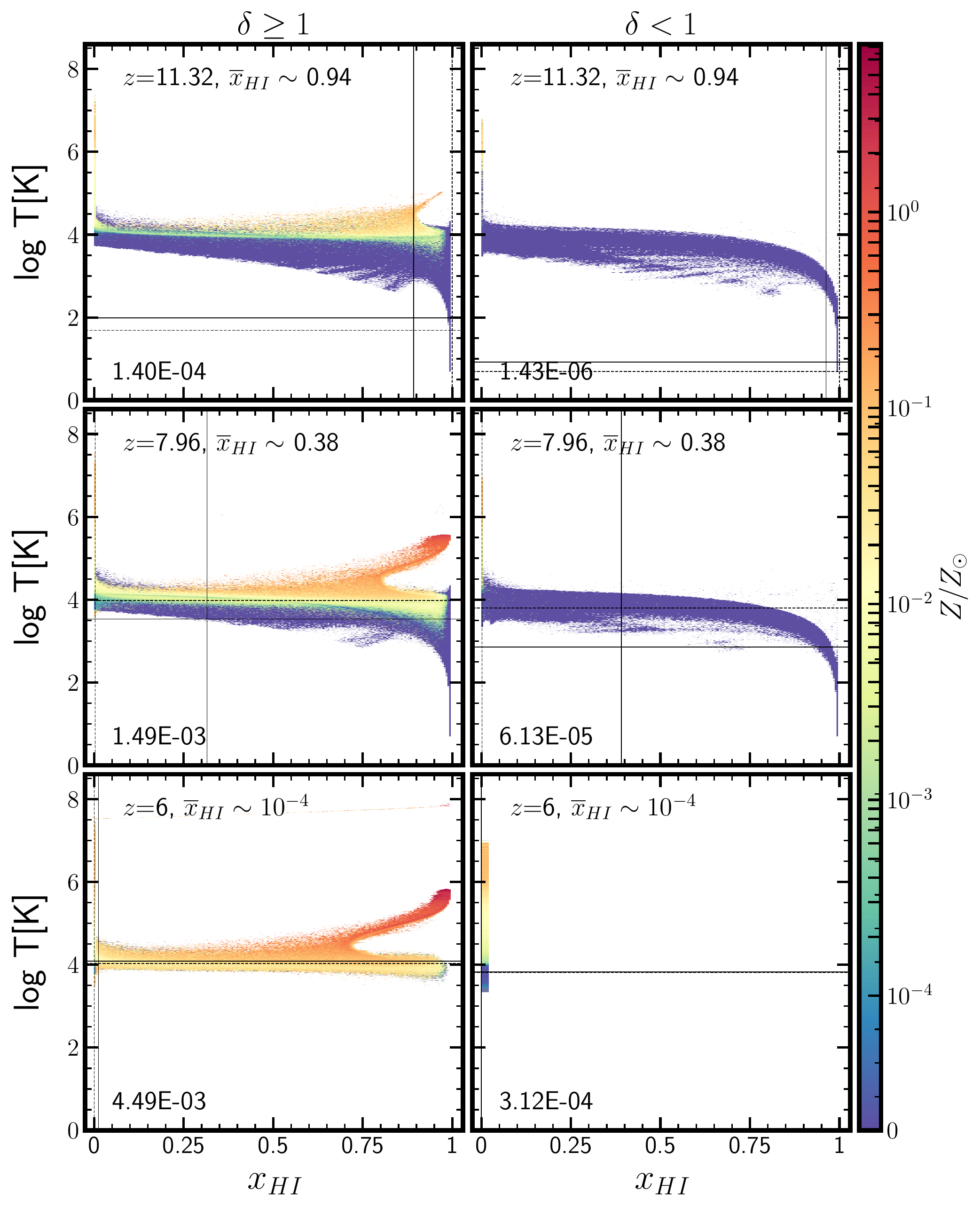}
	\caption{As Figure~\ref{fig:Txhhist} but the color bar indicates the mean gas metallicity relative to the solar metallicity. The numbers in each panel refer to the mass-weighted mean metallicity of the IGM in an overdense/underdense state.}
	\label{fig:TxhZhist}
\end{figure}

In Figure~\ref{fig:Txhhist} we show a scatter plot of gas particle temperature ($T$) as a function of neutral hydrogen fraction ($x_{\rm HI}$) at $z$ = 11.32 (top), 7.96 (middle) and 6 (bottom) in overdense (left) and underdense (right) regions. The numbers in each panel quote the fraction of particles in the given density state.
The physical state of the overdense gas ranges from highly ionized and warm/hot to highly neutral and cold. Most overdense particles are highly neutral and cold at the highest redshift and shift to a highly ionized and warm/hot ($\gtrsim 10^4$~K) state during reionization, although a tail of neutral and warm particles is visible at all redshifts. Underdense particles are also predominantly in a highly neutral and cold state in the early stages and move to a highly ionized and warm/hot state by the end of reionization. However, for underdense gas we do not see a trail of neutral gas at $z$ = 6. 
At all redshifts, we see a small fraction of particles with $T\gg 10^4$~K, which is mostly constituted by gas particles shock heated due to energy from stellar feedback/accretion and hence out of ionization equilibrium. These are seen in various physical states ranging from highly neutral to highly ionized. 

\begin{figure}
	\includegraphics[width=\columnwidth]{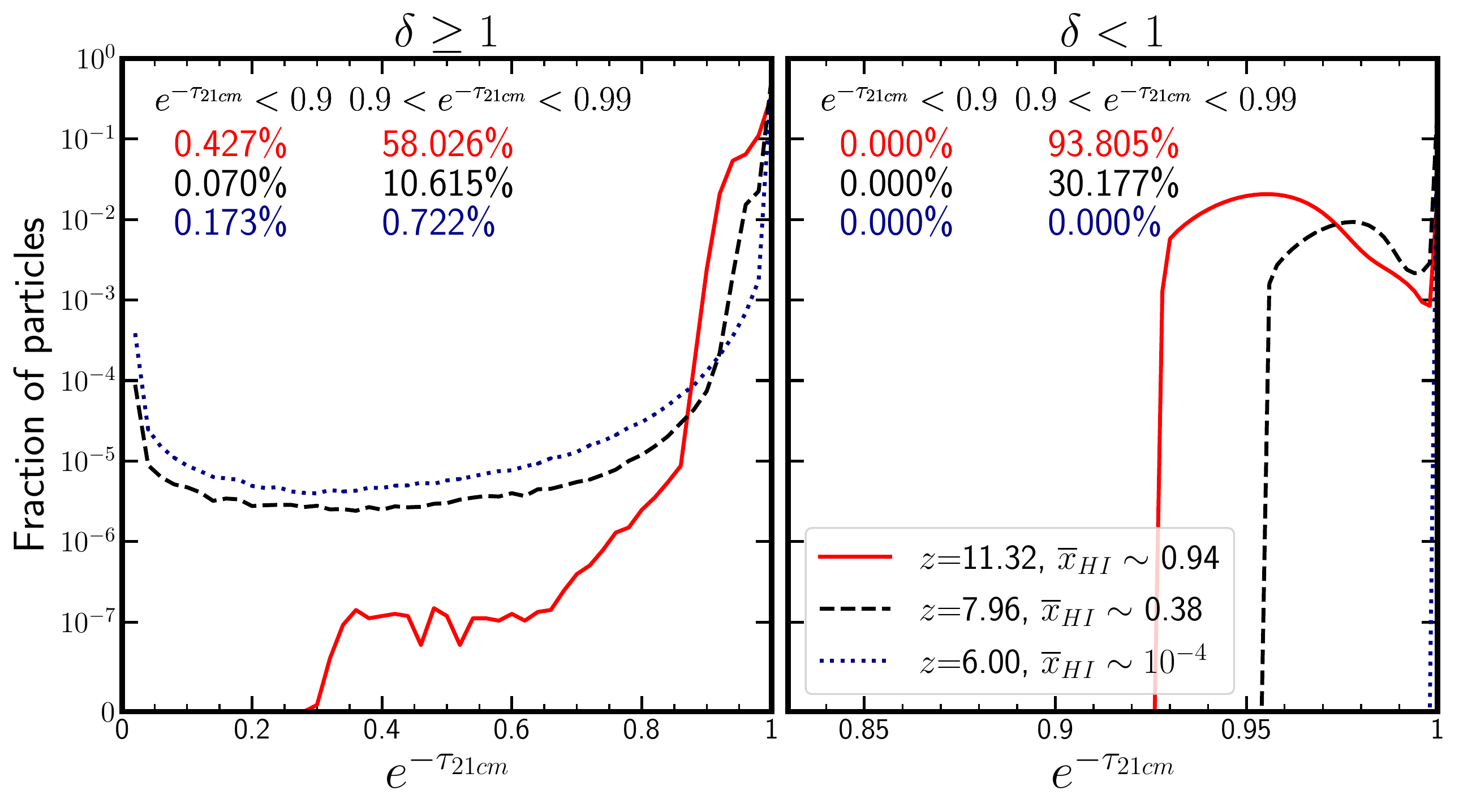}
      	\caption{Distribution of continuum-normalized flux ($e^{-\tau_{21cm}}$) implied by Eq.~\ref{Eq:21cmtau} for the physical properties of each gas particle at $z$ = 11.32 (red solid line), 7.96 (black dashed) and 6.0 (blue dotted) for overdense (left panel) and underdense (right) particles. $\overline{x}_{\rm HI}$ indicates the volume-averaged neutral fraction at each redshift. The percentages in the panels represent the fraction of all particles that are overdense or underdense with $e^{-\tau_{\rm 21cm}} \leq 0.9$ and $0.9 < e^{-\tau_{\rm 21cm}} \leq 0.99$. }
	\label{fig:tauhist}
\end{figure}

\begin{figure}
    \centering
	\includegraphics[width=\columnwidth]{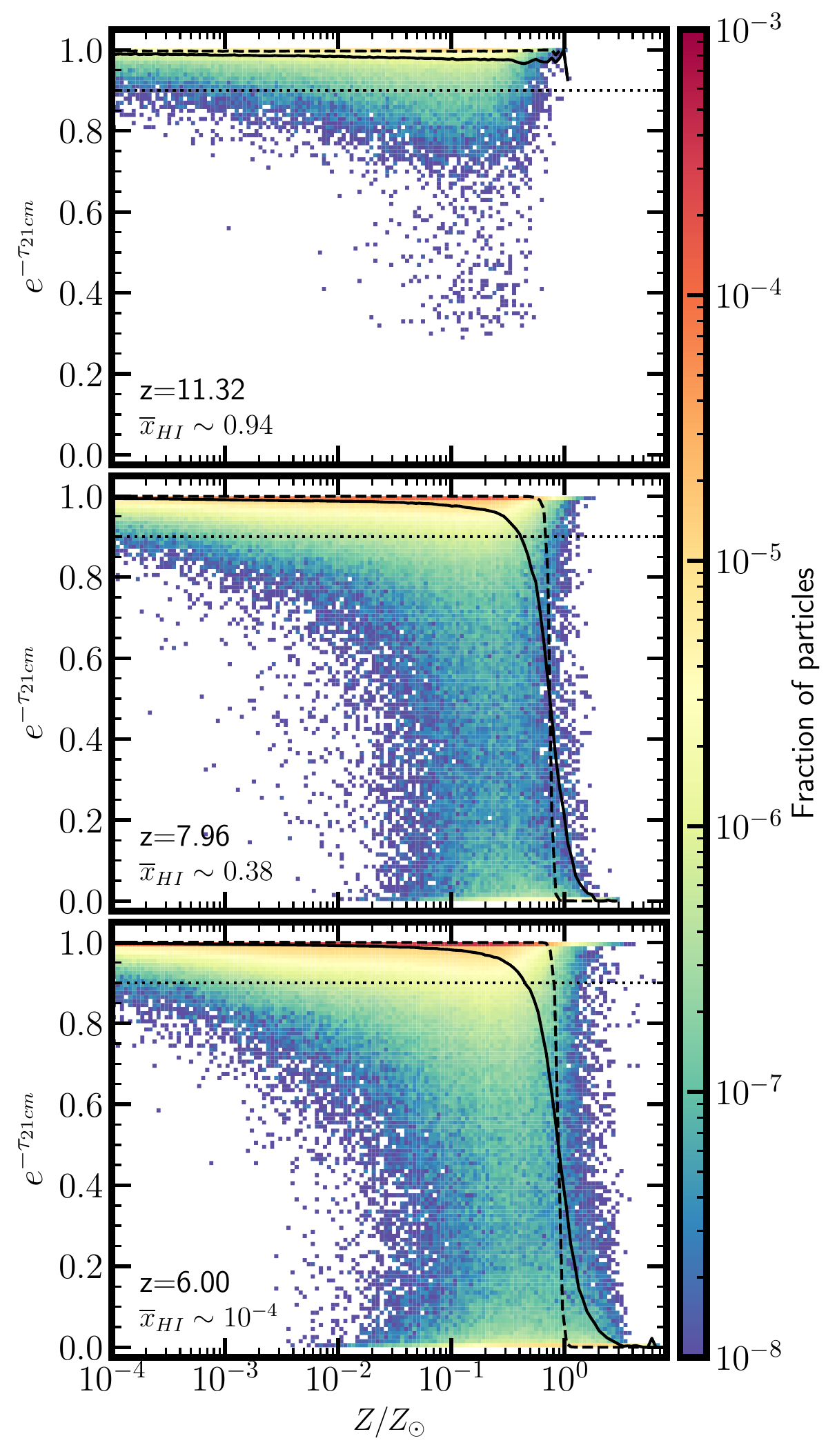}
	\caption{Distribution of continuum-normalized flux ($e^{-\tau_{21cm}}$) implied by Eq.~\ref{Eq:21cmtau} for properties of each gas particle as a function of the metallicity at $z$ = 11.32 (upper panel), 7.96 (middle) and 6 (lower), respectively. $\overline{x}_{\rm HI}$ indicates the volume-weighted neutral fraction at each redshift. The horizontal dotted lines at $e^{-\tau_{\rm 21cm}}$ = 0.9 are drawn to guide the eye. The solid (dashed) line represents the mean (median) $e^{-\tau_{\rm 21cm}}$ for a given metallicity.}
	\label{fig:tauZ}
\end{figure}
In Figure~\ref{fig:TxhZhist} we show a diagram similar to Figure~\ref{fig:Txhhist}, but color coded by the metallicity. As expected, at all redshifts, particles at the highest metallicity are typically warm/hot, neutral and overdense, while the underdense IGM is hardly enriched with metals. The mean metallicity of overdense particles is in fact 1-2 orders of magnitude higher than that of underdense particles at any redshift, as indicated by the numbers in each panel.

As here we are interested in investigating absorption features appearing in 21~cm, in Figure~\ref{fig:tauhist} we show the distribution of the continuum-normalized flux ($e^{-\tau_{\rm 21cm}}$)  implied by Eq.~\ref{Eq:21cmtau} for properties of each gas particle for both underdense and overdense regions. Although we observe the presence of strong absorbers (we refer to strong absorption when $e^{-\tau_{\rm 21cm}}< 0.9$), the vast majority of particles have properties corresponding to weak ($0.9 < e^{-\tau_{\rm 21cm}}< 0.99$) or very weak ($e^{-\tau_{\rm 21cm}} > 0.99$) absorption.
In underdense regions, the absorption arises from the ambient cold IGM. $\tau_{\rm 21cm}$ for $\delta<1$ decreases with redshift, as reionization proceeds in an inside-out fashion and progressively lower gas density gets ionized while the average gas density decreases.
Conversely, in the overdense regions the interplay between the various physical quantities involved is more complex. As seen from the figure, the fraction of strong absorbers declines around $z\approx8$ and increases towards the end of reionization, whereas the fraction of weaker absorbers decreases with the progress of reionization. As seen from the left panel of Figure~\ref{fig:tauhist}, the red line corresponding to z$\approx$11 peaks for the weak absorbers but falls off fast as we move towards the stronger absorbers.

\begin{figure*}
    \centering
	\includegraphics[width=\textwidth]{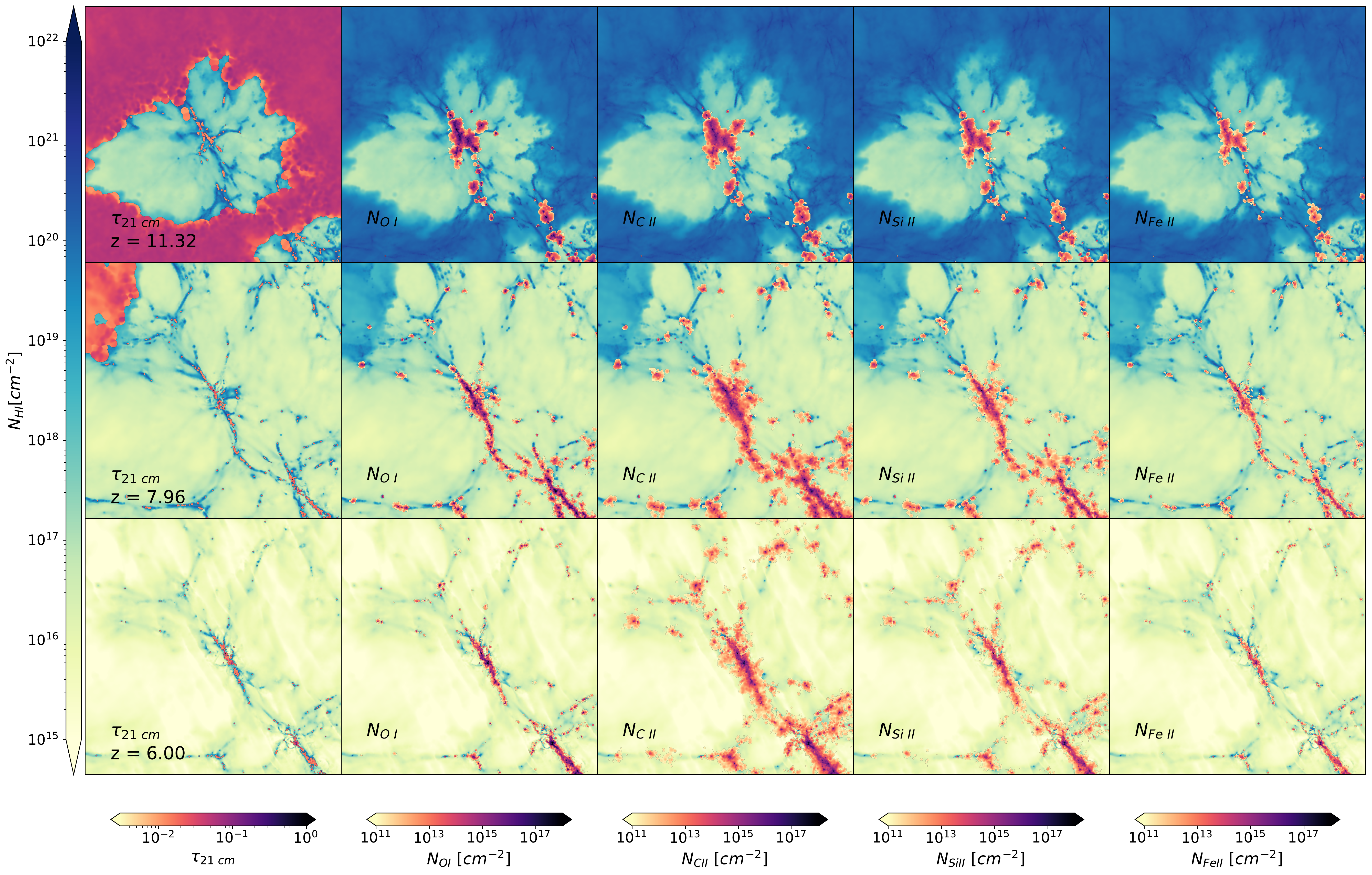}
	\caption{Maps of neutral hydrogen column density overlayed with (from left to right): 21 cm optical depth ($\tau_{\rm 21cm}$), O I, C II, Si II, and Fe II column density at $z$ = 11.32 (top panels), 7.96 (middle), 6.0 (lower). Each projection represents a zoomed in region that is 3~cMpc/$h$ on a side and 1~cMpc/$h$ deep. The region is centered on the particle with the highest metallicity at $z$=7.96.} 
	\label{fig:Slice}
\end{figure*}

Since the overdense gas is the first to be enriched with metals, we are primarily interested in the 21cm absorption that comes from overdense regions. Indeed, the next question we would like to address is whether particles with high 21~cm optical depth are also substantially enriched with metals, so that absorption in both components might be expected.
To investigate this we show, in Figure~\ref{fig:tauZ} the distribution of particles as a function of their $e^{-\tau_{\rm 21cm}}$ (as given by Eq.~\ref{Eq:21cmtau}) and $Z$. Although the vast majority of particles are characterized by low optical depth and metallicity, as long as the overdense gas is self-shielded and not fully ionized, we observe particles with $e^{-\tau_{\rm 21cm}}<0.9$ and metallicity $Z>10^{-3} Z_{\odot}$. From the solid (dashed) line showing the mean (median) $e^{-\tau_{\rm 21cm}}$ values with respect to the metallicity, we can see that the strongest absorption in 21 cm comes from particles with the largest metallicities.

In Figure~\ref{fig:Slice} we show a 1 cMpc/$h$ thick slice centered on the particle with the highest metallicity ($Z$ = 3.024 $Z_\odot$) at $z$ = 7.96. The columns show neutral hydrogen column density overlayed with 21 cm optical depth, {\tt O I, C II, Si II} and {\tt Fe II} column density from left to right respectively. The evolution of the absorbers is well illustrated in this figure. The 21 cm signal evolves from showing absorption predominantly from the ambient IGM to being confined within the densest, self-shielded regions by the end of reionization. 

\subsection{Absorption Spectra}
\label{sec:AbsSpec}
\begin{figure*}
    \centering
    \includegraphics[width=\textwidth]{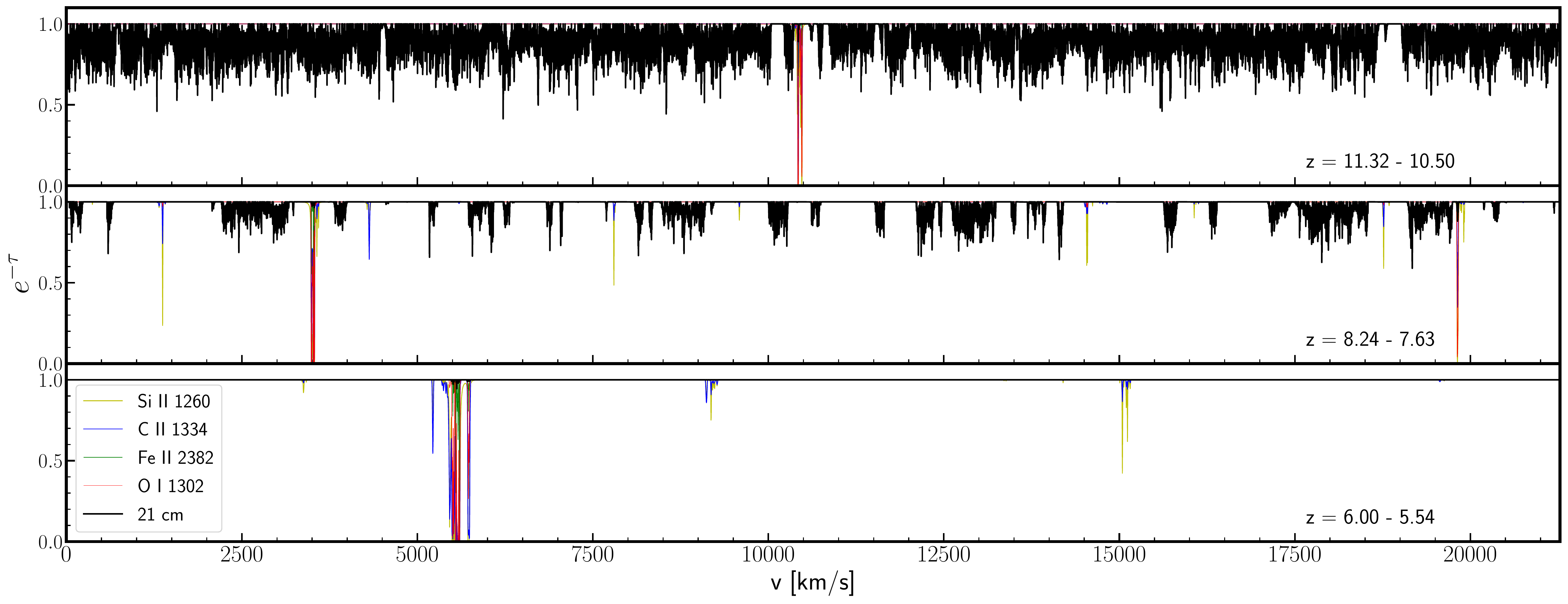}
    \caption{Example of sightlines between $z$ = 11.32 - 10.50 (top panels), $z$ = 8.24 - 7.63 (middle panel) and $z$ = 6.00 - 5.54 (bottom panel) showing the 21 cm forest absorption spectra (black) overlayed with metal spectra for O I 1302\AA \ (red), C II 1332\AA  \ (blue), Si II 1260\AA \ (yellow) and Fe II 2382\AA \ (green). The spectra can be converted from the velocity space $v$ to the observed wavelength using $\lambda = \lambda_0 (1+z)(1+v/c)$, where $\lambda_0$ is the rest frame wavelength for the transition.}
	\label{fig:los_example}
\end{figure*}
To calculate both metal and 21~cm synthetic absorption spectra, we use the package {\tt TRIDENT} \citep{Trident2017} which is a python based tool integrated with {\tt yt} \citep{YT2011}.
In brief, {\tt TRIDENT} evaluates the ionization balance of each gas particle along a line of sight (LOS) using tables pre-computed with the photoionization package $\mathrm{CLOUDY}$ \citep{cloudy} assuming that the gas is exposed to the \cite{HM12} metagalactic UV background\footnote{For neutral gas, the ion balance calculated using the UV background shows that neutral (for Oxygen) and singly ionized states (for Carbon, Silicon and Iron) dominate the ionization states for the metals of interest. We also highlight that, although the buildup of a UV background in the simulation is not strictly consistent with the assumption of a Haardt \& Madau one, we have verified that adopting different shapes and intensities for the UVB does not change our conclusions, which we then consider robust.} (note that the neutral hydrogen fraction is instead evaluated directly in the Aurora simulations). We also set the temperature of starforming gas to $10^4$ K during ion balance calculation for metals, since the equation of state for this high-density gas does not reflect the temperatures we expect from the ISM. This treatment of gas on the equation of state has a negligible effect on absorber statistic.
For metal species, the absorption feature produced by each gas particle is represented by a Voigt profile with a line center at velocity $v$ and a doppler parameter $b$ specified by the temperature of the gas particle. Each spectrum is binned in the velocity space and {\tt TRIDENT} returns the optical depth and normalized flux as a function of velocity along the LOS. The spectra can be converted from the velocity space $v$ to the observed wavelength using $\lambda = \lambda_0 (1+z)(1+v/c)$, where $\lambda_0$ is the rest frame wavelength for the transition. The code has been modified to use Eq.~\ref{Eq:21cmtau} to evaluate also the 21~cm optical depth.

The lines of sight are projected at random orientations, but they extend for a fixed redshift interval, $dz$, which  is set such that the metal transition of interest remains visible (i.e. redward of the Lyman alpha forest) if we assume that a quasar is located at the end of each LOS \citep{furlanetto2003metalabs}. Since we are primarily interested in the O I 1302 \AA~line, we use this to set the $dz$ for our LOS. 

More specifically, we evaluate 100 random LOSs in the intervals $\Delta z=11.32-10.50$, 8.24-7.63 and 6.00-5.54, respectively. We are keenly interested in the midpoint of reionization since the interplay between metal enrichment and 21 cm forest evolution gives us the most interesting scenarios at this epoch. Additionally, the observational counterparts of the absorption lines studied here are most likely to be detected at these intermediate redshifts with the capabilities of ELT and SKA (see sec.~\ref{sec:Observations}). 
Although the observation of such a large number of lines of sight at $z\ge6$ is unlikely at the moment, these can still provide us with useful insight into the statistical significance of high-$z$ absorption studies, which will become relevant for upcoming instruments. 

In Figure~\ref{fig:los_example} we show ideal noiseless absorption spectra (i.e. without instrumental characteristics) along randomly chosen sightlines at different redshifts, as described previously. 
We can clearly see the redshift evolution of the 21 cm forest as the fraction of photoionized warm gas ($T\approx$ $10^4$~K) increases and the absorption decreases accordingly. Indeed, the number and size of the transmission windows become larger towards lower redshifts, as reionization proceeds. Although the redshift evolution of metals is not evident from a single sightline, as gas gets enriched with metals, the incidence of metal line absorbers increases drastically between $z$ = 11.32, where only about 29\% of the sightlines show metal-line absorption, and $z$ = 6, where this percentage has increased to 96\%.

\subsection{Aligned absorbers}
\label{sec:AlignedAbs}
In this study, we focus on the special case where 21~cm and metal lines occur as aligned absorbers. We define aligned absorbers as follows: first, we overlay the spectra of all metal ions of interest along a sightline in velocity space, then find windows in velocity space (these windows span a few tens to a few hundred km/s in velocity space) that have absorption from one or more metal ions of interest. If two neighbouring windows lie within 100 km/s of one another, we combine them into one. We then overlay the 21 cm spectrum to check if the windows show also 21~cm absorption. When this happens we call it an aligned absorbing window. Depending on the total window equivalent width and the degree of 21 cm absorption within a given window, we classify an aligned absorbing window as Aligned and Cospatial Absorber (ACA) or Aligned but Not Cospatial Absorber (ANCA) (see below for their definition). 

The race between reionization of neutral gas and metal enrichment gives rise to a very wide range of absorption scenarios. Sightlines at each redshift provide us with unique information and an overall picture of the complex redshift evolution of aligned absorbers. We examine each redshift individually.

\subsubsection{Redshift 11.32, $\overline{x}_{\rm HI}$ $\approx$ 0.94}

\begin{figure*}
    \centering
	\includegraphics[width=\textwidth]{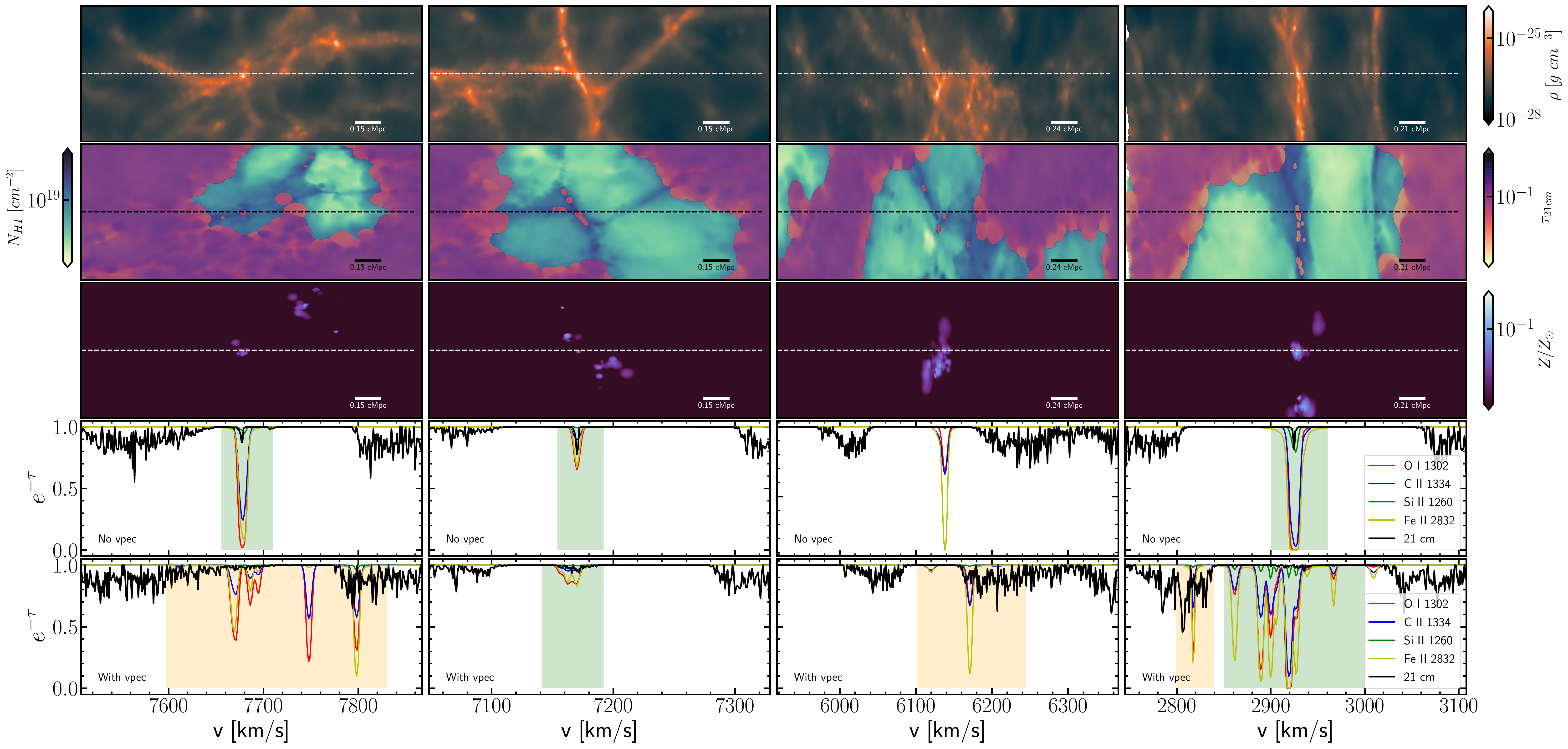}
	\caption{Zooms into four different absorbing windows extracted randomly from the sample of sightlines at $z$= 11.32. From top to bottom the rows refer to: gas density, neutral hydrogen column density overlayed with the 21 cm optical depth, gas metallicity, absorption spectra with $v_{\rm pec}$ = 0 and absorption spectra with $v_{\rm pec}$. Green shaded regions highlight aligned and cospatial absorbers (i.e. arising from the same pocket of gas), while the orange shaded regions highlight aligned but not cospatial absorbers (i.e. they appear aligned due to displacements induced by peculiar velocity effects).}
	\label{fig:aligned_40}
\end{figure*}

At these high redshifts, metal absorbers are predominantly seen in windows where there is complete transmission in 21~cm. Indeed, stars driving the winds and/or SN explosions that enrich gas with metals also ionize and heat it, creating transmission in 21~cm. However, as seen from Figure~\ref{fig:aligned_40}, within such transmission regions, isolated small scale 21 cm absorbers can exist. The key feature of aligned absorbers at these redshifts is that they are embedded in the thick 21 cm forest (see also the upper panel of Figure~\ref{fig:los_example}). 

We note here that peculiar velocities play a very important role in the interpretation of aligned absorbers, as they can displace metal lines and make them appear to be aligned with 21~cm lines, while they might instead arise from different locations. 
This is shown in columns 1, 3 and 4 in Figure~\ref{fig:aligned_40}, where metal lines appear to be aligned with the 21~cm forest in the cold and underdense IGM because of the peculiar velocity effects, while in reality they arise from different gas. Specifically, we note the absorbing window in Column 3, where the metal absorber has no alignment at all (metal absorption in a transmission window) but appears aligned due to peculiar velocity effects. 
From now on we call these aligned but not cospatial absorbers (ANCAs). They are characterized by high 21 cm equivalent widths (EW $>$ 1 km/s) and strong 21 cm absorption ($e^{-\tau_{21cm}} < 0.9$), with a mean (median) fraction of pixels within the window with 21 cm absorption ($e^{-\tau_{21cm}} < 0.99$) of $\approx$ 35\% (27\%). On the other hand, a situation as shown in column 2 in Figure~\ref{fig:aligned_40} is the case where aligned absorption originates from the same underlying gas. From now on we call these aligned and cospatial absorbers (ACAs). They are characterized by modest 21 cm equivalent widths (EW $\le$ 1 km/s), and weak 21 cm absorption ($0.9 < e^{-\tau_{21cm}} < 0.99$) which is localized to a small number of pixels within the absorbing window, with a mean (median) fraction of pixels with 21 cm absorption ($e^{-\tau_{21cm}} < 0.99$) of $\approx$ 5\% (2\%).
We find that ANCAs are predominant at higher redshifts, when the 21 cm forest is abundant. 
Although ANCAs are not useful to infer the properties of the underlying gas since the alignment is only apparent and the absorption originates from different underlying gas, they can provide information on the gas peculiar velocity based on the relative position of the absorbing windows and the nearby transmission gaps in the 21 cm forest.
In our sample, we find 2 (41) sightlines (absorbers) with metals and 23 (31) sightlines (absorbers) with aligned absorbing windows. Out of the 31 aligned absorbing windows detected, 22 are ANCAs and 4 are ACAs which show strong 21 cm absorption (EW > 0.1 km/s) within the absorbing window and 5 ACAs which show weak 21 cm absorption (0.01 km/s < EW < 0.1 km/s). The incidence of ACAs is low at this redshift, which makes observability difficult without a large sample of sightlines. However, ANCAs are more likely to be detected.

\subsubsection{Redshift 8.24, $\overline{x}_{\rm HI}$ $\approx$ 0.48}
\begin{figure*}
    \centering
	\includegraphics[width=\textwidth]{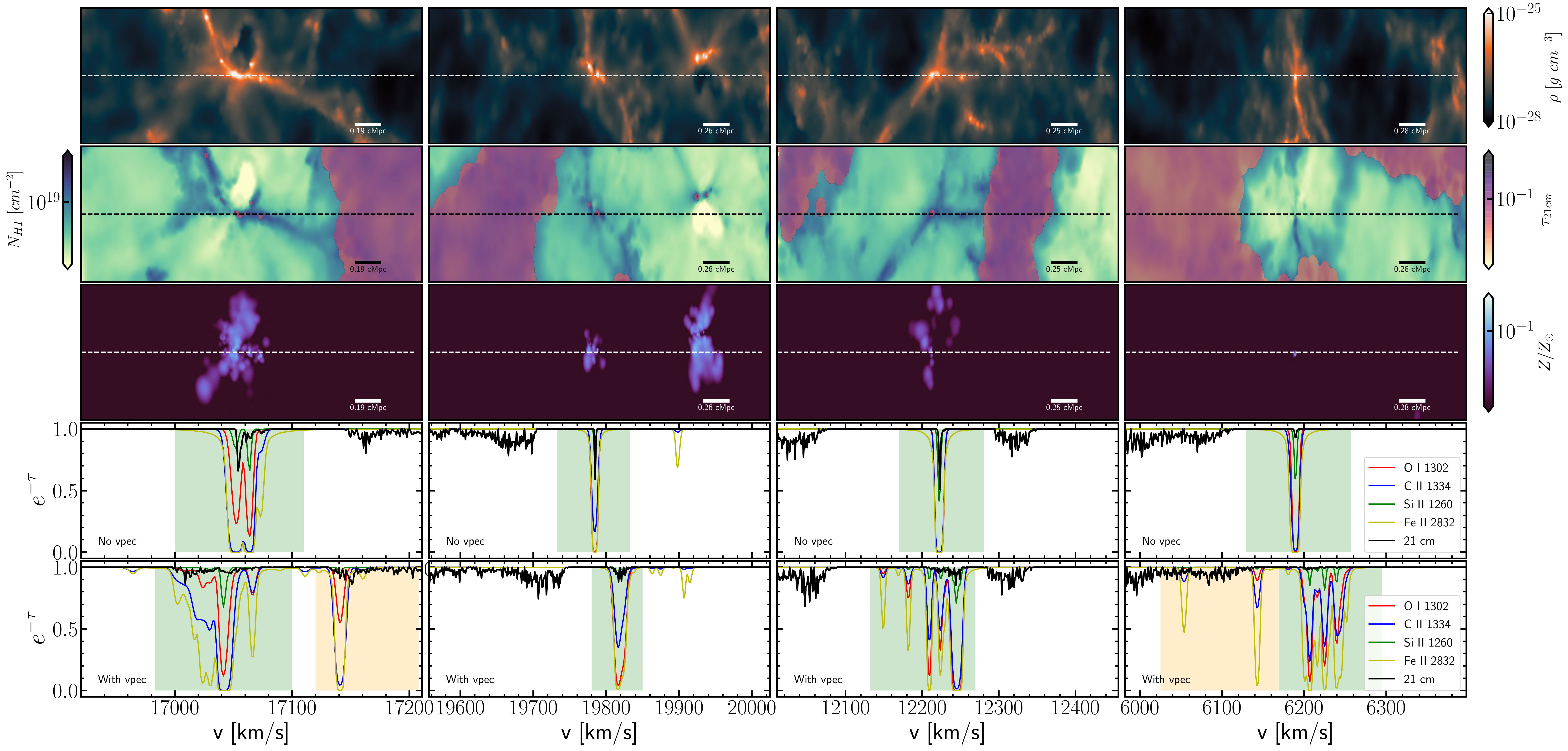}
	\caption{As Figure~\ref{fig:aligned_40} but at $z$ = 8.24.}
	\label{fig:aligned_51}
\end{figure*}
As reionization proceeds, the HI column density and associated absorption decrease (as clearly seen in Figure~\ref{fig:aligned_51}), and more transmission gaps open up in the 21 cm forest. At the same time, metals are more widely spread. Hence,  most metal absorbers are located in the 21~cm transmission gaps. However, there is still enough of a 21 cm forest that comes from the ambient cold and underdense IGM to produce both ANCAs and ACAs. We find 87 (227) sightlines (absorbers) with metals and 55 (79) sightlines (absorbers) with aligned absorbing windows. Out of the 79 aligned absorbing windows, 11 are ANCAs, 29 are ACAs with strong (EW > 0.1 km/s) and 39 ACAs with weak (0.01 km/s < EW < 0.1 km/s) 21 cm absorption. Observing the latter would be very challenging also with an instrument like the SKA. The ACAs remain an excellent probe of the densest neutral structures as demonstrated in Figure~\ref{fig:aligned_51}. 

From a comprehensive analysis of all the absorbers, we find a range of scenarios in which they can emerge. Some occur in isolation within transmission windows, whereas others are seen at the edge of a dense patch of 21 cm forest. This can provide us a great deal of information about the underlying structure, surrounding gas environment and peculiar velocity. At these intermediate redshifts, the incidence of ACAs is high, making this redshift range the most observationally viable, as for a fixed number of sightlines the chance of detection of one of such absorbers is higher.
\begin{figure*}
    \centering
	\includegraphics[width=\textwidth]{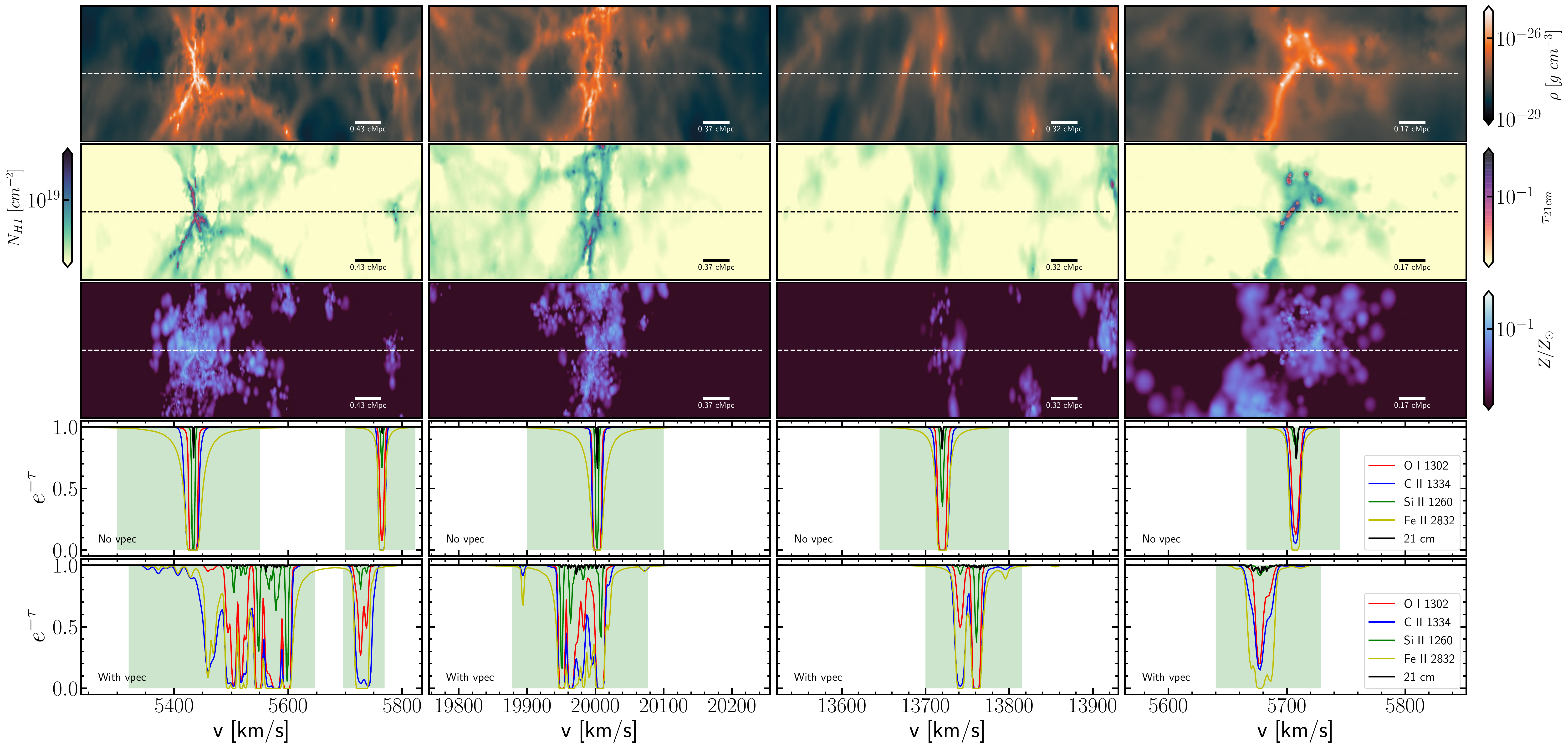}
	\caption{As Figure~\ref{fig:aligned_40} but at $z$ = 6.00.}
	\label{fig:aligned_60}
\end{figure*}

\begin{figure*}
    \centering
	\includegraphics[width=\textwidth]{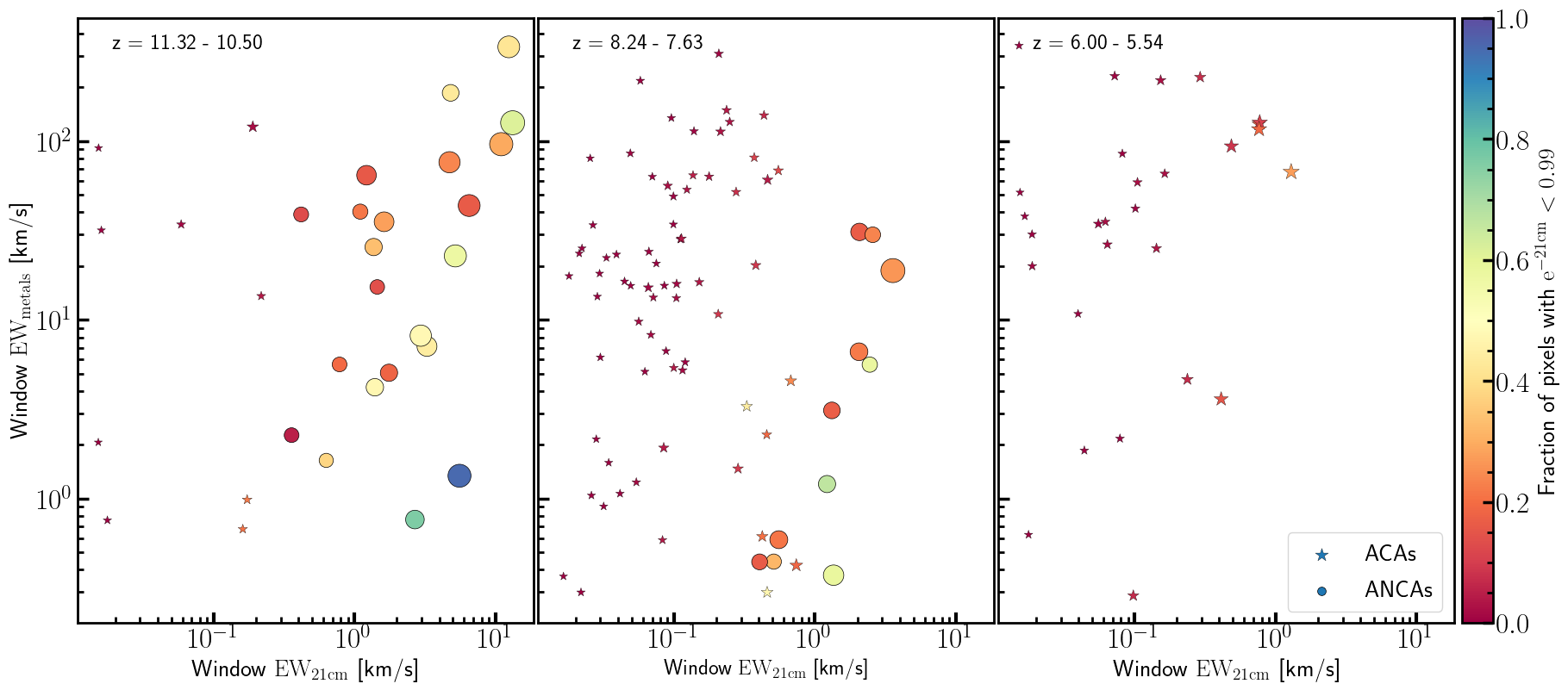}
	\caption{Rest frame total window equivalent width for metals vs rest frame total window equivalent width for 21 cm for sightlines in the range $z$ = 11.32 - 10.50 (left panel), $z$ = 8.24 - 7.63 (middle) and $z$ = 6.00 - 5.54 (right). Stars and circles indicate ACAs and ANCAs, respectively. The colorbar shows the fraction of pixels in the absorbing window with 21 cm absorbers ($e^{-\tau_{21cm}}$ < 0.99). The sizes of the symbols indicate the strength of the strongest 21 cm absorbing pixel  within a given window, ranging from $e^{-\tau_{21cm}}$ = 0.98 for the smallest symbols to $e^{-\tau_{21cm}}$ = 0.7 for the largest. }
	\label{fig:EWdist_all}
\end{figure*}

\subsubsection{Redshift 6.00, $\overline{x}_{\rm HI}$ $\approx$ $10^{-4}$}
By the end of reionization, the neutral gas is predominantly confined to small scale pockets of gas within the large-scale structure, as shown in Figure~\ref{fig:aligned_60}, with the 21 cm forest being overall weak and confined to very small regions. We find 96 (355) sightlines (absorbers) with metals and 27 (28) sightlines (absorbers) with aligned absorbing windows. The incidence of such absorbers decreases significantly by the end of reionization owing to the declining 21 cm optical depth. Out of the 28 absorbing windows, 0 are ANCAs, 13 ACAs which show strong 21 cm absorption (EW > 0.1 km/s) and 15 ACAs with weak absorption (0.01 < EW < 0.1 km/s). Because the 21 cm forest is minimal and confined to a very small region within the absorbing windows, we do not see any ANCAs at this redshift, while ACAs remain excellent tracers of the underlying structure, as the neutral gas that leads to 21 cm absorption is predominantly located in the densest structures. 
As seen from the 4th and 5th rows of Figure~\ref{fig:aligned_60}, peculiar velocities can affect absorption windows in many different ways. 

Figure~\ref{fig:EWdist_all} shows the complete picture of incidence and evolution of both ACAs and ANCAs throughout reionization. We can see that the ACAs and ANCAs separate themselves into distinct parts of the equivalent phase space. ACAs have weak 21 cm EWs (EW < 1 km/s) while ANCAs have strong 21 cm EWs (EW > 1 km/s). ACAs show a low fraction of pixels within the window with 21 cm absorption (~5\%) whereas ANCAs typically have a high fraction of pixels within a window with 21 cm absorption (~35\%). The sizes of the markers in the plot show the strength of the deepest 21 cm absorber within each absorber. ANCAs have the largest markers showing absorbers ($e^{-\tau_{21cm}} < 0.9$), while ACAs with the smaller markers show modest-weak absorption ($0.9 < e^{-\tau_{21cm}} < 0.99$). At early times, due to the presence of a thick 21 cm forest from the cold underdense IGM, the ANCAs dominate, with only a handful of ACAs. Around midway through reionization, as the 21 cm forest thins out, the incidence of ANCAs reduces while ACAs become abundant. At the end of reionization we do not see any ANCAs but ACAs are still abundant. In summary, the incidence of ANCAs decreases with decreasing redshift, while for ACAs it peaks around the midpoint of reionization.

\section{Observability}
\label{sec:Observations}
In the following, we will assess the prospects of detecting the redshifted {\tt O I} metal absorption in the near-infrared using the upcoming ArmazoNes high Dispersion Echelle Spectrograph
\citep[ANDES, previously referred to as HIRES;][]{Marconi21} on the ELT, and the redshifted 21 cm absorption using SKA1-LOW. Here we focus on {\tt O I} because it is the best tracer of neutral hydrogen and because, due to its weak oscillator strength, {\tt O I} absorption lines are typically unsaturated. For this purpose, as a background source we assume a $z\approx 8$ quasar with a rest-frame UV continuum corresponding to an AB magnitude at 1450 \AA \; of $m_{\rm AB}=21.0$~mag. For comparison, both the most distant quasar ($z\approx 7.64$; \citealt{Wang21}) and the most distant radio-loud quasar ($z\approx 6.82$; \citealt{Banados21}) currently known have $m_{\rm AB} \approx 21$~mag in the near-infrared $J$-band. Quasars as bright as $m_{\rm AB} \approx 21$~mag are admittedly expected to be rare at $z\approx 8$, but the modelling by \citet{Barnett19} suggests that a few such objects may be detected by the upcoming Euclid wide survey. For our reference SKA simulation, we assume a radio continuum flux of 10~mJy at 150~MHz, and a spectral index $\alpha$ = -1.31, following the observations of the most distant radio loud quasar ($z\approx 6.82$; \citealt{Banados21}). 

\subsection{21 cm forest observations with SKA1-LOW}
\begin{figure*}
    \centering
	\includegraphics[width=\textwidth]{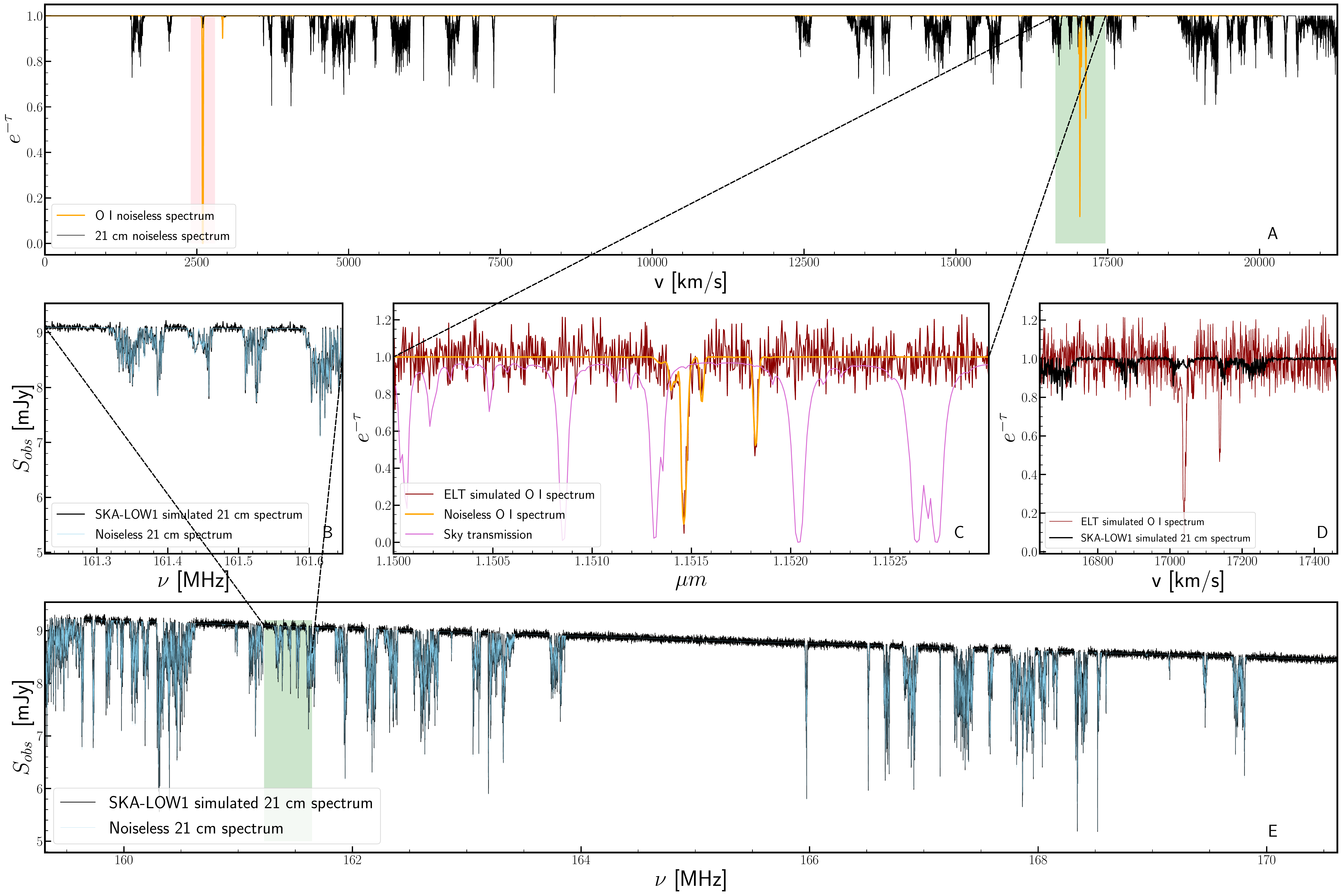}
	\caption{Example of synthetic spectrum and mock observations for the ELT-ANDES instrument and SKA1-LOW.  {\it{Panel A}}: Synthetic (noiseless) spectra at z $\approx$ 8 showing the 21~cm forest (black lines) and the {\tt O I} 1302 absorption spectra (orange). Shaded regions (pink and green) highlight the location of ACAs, although their extent does not coincide with that of the corresponding absorber. The green shaded regions correspond to the same absorber in panels A and E. {\it{Panel B}}: Zoom into the portion of the 21 cm forest spectrum with the ACA shown in Panels A and E. The blue line is the noiseless spectrum ($e^{-\tau_{21cm}}$), while the black one is the simulated spectrum ($S_{obs} = S_0 e^{-\tau_{21cm}}$) for a background radio loud source observed with SKA1-LOW, assuming background source flux ($S_0$) of 10~mJy at 150~MHz, a channel width of 10~kHz and an observing time of 1000~h. {\it{Panel C}}: Zoom into the portion of the {\tt O I} spectrum of Panel A with the ACA. The orange line shows the noiseless spectrum, while the red line shows the simulated spectrum as observed by the ELT-ANDES instrument assuming an exposure time of 10~h, 2 readouts per hour and spectral resolution $R=1\times 10^5$ for a $m_{\rm AB} = 21.0$ mag source. The purple line shows sky transmission at these wavelengths. {\it{Panel D}}: Simulated normalized spectra as observed by the ELT and SKA-LOW1 for {\tt O I} and 21 cm overlayed in velocity space showing the green shaded region from Panel A. This clearly shows the alignment of the absorption features. {\it{Panel E}}: Simulated 21 cm forest for a radio loud source with continuum-flux ($S_0$) of 10~mJy (at 150~MHz) with a power law index of -1.31. The blue line shows the noiseless spectrum (as in panel A) multiplied with the radio continuum of the background source, while the black line shows simulated spectrum as observed by SKA-LOW1.}
	\label{fig:ELTSKAsim}
\end{figure*}

Since SKA1-LOW is still in the design phase, we simulate observations using available telescope configurations and a sky model at frequencies of interest. We use the OSKAR (\citealt{mort2010oskar}) simulator to generate the visibility data from the interferometer. Phase-1 of SKA1-LOW has a baseline design of 512 stations that each contain 256 dual-polarization antenna elements with a maximum baseline of 65~km. Our simulations use the SKA1-low V5 configuration which consists of 512 stations. This gives a telescope model comprising the entire array over the full range of baselines. We calculated the Stokes-I amplitudes assuming a sky model where our radio loud source of interest is the only source in the sky. Noise is added using the following equations. The RMS noise in a single polarisation of a receiver is:
\begin{equation}
\sigma_{S}^{sin} = \frac{2 k_{B} T_{sys} \eta}{A_{eff}\sqrt{2 \tau \Delta \nu}}\,
\end{equation}
with $T_{sys}$ the system temperature, $\eta$ the efficiency, $A_{eff}$ the effective area, $\Delta\nu$ the channel bandwidth, and $\tau$ the observing time.
For measurements comprised of combinations of two independent receiver polarisations, such as those expressed as Stokes parameters, this will be reduced by a factor of $1/\sqrt{2}$, and therefore:
\begin{equation}
\sigma_{S} = \frac{\sigma_{S}^{sin}}{\sqrt{2}} = \
\frac{k_{B} T_{sys} \eta}{A_{eff}\sqrt{\tau \Delta \nu}}\ .
\end{equation}
The RMS noise on a baseline ($p, q$) is given by $\sigma_{S}^{p, q} = \sqrt{\sigma_{S}^{p}\sigma_{S}^{q}}$, under the assumption that the telescope is composed of stations of equal sensitivity, then $\sigma_{S}^{p, q} \equiv \sigma_{S}$. The telescope sensitivity values are obtained for each of our frequency cubes and noise is added appropriately. The ratios of effective area and system temperature are given in Table~\ref{tab:Tsys_Aeff} \citep{braun2019anticipated}.

\begin{table}
\centering
\begin{tabular}{|r|r|r|}
Frequency (MHz) & SKA1-Low Station & SKA1-Low Array \\
& $ \frac{A_{\rm eff}}{T_{\rm sys}} $[m$^2$/K] & $ \frac{A_{\rm eff}}{T_{\rm sys}} $[m$^2$/K]\\
\hline
50   &   0.102  &   52.2      \\
70   &   0.420  &   214.8     \\
100  &   0.862  &   441.3     \\
110  &   0.885  &   453.1     \\
120  &   0.924  &   472.9     \\
150  &   1.119  &   572.8     \\
200  &   1.223  &   626.4     \\
300  &   1.289  &   660.0     \\
350  &   1.156  &   591.8    
\end{tabular}
\caption{Individual SKA1-Low stations and combined SKA1-Low array natural sensitivities as a function of frequency, averaged over solid angles within 45 degrees of the zenith (the values are taken from \citealt{braun2019anticipated})}
\label{tab:Tsys_Aeff}
\end{table}
For our sightlines, we simulate image cubes starting at 202.91 MHz, 153.72 MHz and 115.29 MHz with $\approx$11 MHz, $\approx$11 MHz, $\approx$8 MHz bandwidths. The simulation was run to mimic a 1000~h observation for a radio source with continuum flux of 10~mJy at 150~MHz. For the sake of computational speed, the visibility coordinates were generated for a 10 h observation, while the noise was added assuming a 1000 h observation. This procedure is valid under the assumption that the exact same 10~h observation was run over 100 nights. The background radio source  was assumed to have a power law and flux as described previously. Following the visibility calculation, the image cube was assembled and the sightline spectrum was extracted.

In Figure~\ref{fig:ELTSKAsim}-A,B,E we show the noiseless synthetic spectra and simulated spectra at $z$ $\approx$ 8. Figure~\ref{fig:ELTSKAsim}-A (black line) shows the noiseless spectrum as calculated from the sightline. Figure~\ref{fig:ELTSKAsim}-E shows the noiseless (blue line) and the full simulated source (black line) spectrum as seen by the SKA. For the 1000h observation,  we can clearly see the strong 21 cm features in the spectrum. Figure~\ref{fig:ELTSKAsim}-B shows a region in the spectrum where ACAs are detected. The zoomed in region clearly shows how the simulated SKA-LOW1 spectrum is very close to the noisless spectrum and the strong absorbers are very well resolved.

\subsection{Metal detection with ELT/ANDES}
In Figure~\ref{fig:ELTSKAsim}-A (orange line) shows the synthetic noiseless {\tt O I} spectrum. Figure~\ref{fig:ELTSKAsim}-C, D, we show the {\tt O I} features (ACA) in a mock spectrum based on S/N ratios predicted using the ELT/ANDES (HIRES) exposure time calculator\footnote{\url{https://elt.eso.org/instrument/ANDES/}} v.2.0. Here, we assume an exposure time of 10 h, two readouts per hour and a spectral resolution of $R=1\times 10^5$ for a $m_{\rm AB} = 21.0$ mag source, giving S/N$\approx 20$ per resolution element at the continuum limit. To produce the mock spectrum, the noise-free spectrum has been convolved with a Gaussian line spread function and resampled at 3 pixels per resolution element before adding the noise according to the flux-dependent S/N. Significant detections of at least three {\tt O I} absorption features are seen from Figure~\ref{fig:ELTSKAsim}-C. With an absorber redshifts of $z\approx 7.84$, the {\tt O I} features appear at observed wavelengths of $\approx 1.15 \micron$, in a region between the photometric $Y$ and $J$ bands subject to strong telluric absorption as shown in Figure~\ref{fig:ELTSKAsim}-C (purple line). For this particular absorption complex, the Cerro Paranal Advanced Sky Model indicates that the relevant {\tt O I} features will not be strongly affected by sky absorption lines at the spectral resolution of ELT/ANDES. The low-wavelength wing of the {\tt O I} absorption component at $\approx 11514$ \AA{} may, however, be partially affected by an upper-atmosphere airglow line.

\section{Conclusions}
\label{sec:conclusions}

In this paper we study the correlation between 21 cm and low ionization state metal absorbers using the high-resolution radiation-hydrodynamical simulation of the EoR from the Aurora project. We project sightlines to calculate absorption spectra for {\tt O I, C II, Si II, Fe II} and the 21 cm forest, overlay metal and 21 cm spectra and characterize absorbing windows to find correlations between the metal and 21 cm spectra. We then create mock observations for SKA1-LOW and ELT/ANDES to evaluate the observational feasibility of such a study. The analysis is done at $z\approx$11, 8 and 6. Our key results are summarized as follows: 
\begin{itemize}
    \item We show the redshift evolution of the physical state of gas through reionization. We find overdense gas particles that are metal rich but neutral all the way to the end of reionization. We show that metal rich particles with metallicities $Z/Z_{\odot} > 10^{-3}$ can also have high ($\tau_{21cm} > 0.01$) 21 cm optical depths down to redshift 6. By the end of reionization the overdense phase of gas is responsible for the majority of the 21 cm absorbers, as opposed to the early stages when they mostly arise in the cold underdense IGM.
    \item Aligned and cospatial absorbers (ACAs) (i.e. both metal and 21~cm lines arise from the same location) are excellent tracers of the densest gas. Aligned but not cospatial absorbers (ANCAs) (i.e. metal and 21~cm absorption appear aligned due to displacements induced by peculiar velocity effects) originate instead preferentially in the ambient underdense IGM gas. The two kinds of absorbers can provide different information about underlying gas, environment and peculiar velocities. 
    \item At all redshifts we find windows with absorption in both metals and 21~cm, but their incidence shows a strong redshift dependence, as dictated by the interplay between the evolution of the 21 cm forest and the metal enrichment of the gas. The incidence of ANCAs decreases steadily with decreasing redshift whereas the incidence of ACAs is maximum at z $\approx$ 8. While ANCAs disappear towards the end of reionization because of the lack of an ambient 21 cm forest, ACAs survive (albeit with a reduced strength and number) as they arise from  overdense self-shielded pockets of neutral gas.
    \item To investigate to observability of such absorbers, we assume a background quasar at $z\approx 8$ with a rest-frame UV continuum corresponding to an AB magnitude at 1450 \AA \; of $m_{\rm AB}=21.0$~mag and a radio continuum flux of 10 mJy. Thin unsaturated absorption lines of {\tt O I} should be detectable with the ELT/ANDES instrument with a spectral resolution of R = $1\times 10^5$ and 10~h exposure time. 
    \item The strong 21 cm absorbers (EW > 0.1 km/s) should be readily detectable with the SKA for a 1000~h observation, while the weak absorbers (EW < 0.1 km/s) will be challenging to detect reliably. 
\end{itemize}
The main challenge to such a study is the need for high redshift quasars that are simultaneously radio {\it{and}} infrared loud. So far, despite the several tens of quasars detected at $z>6$ only a handful are radio loud  \citep[e.g][]{zeimann2011discovery,reed2015j0454,li2020scuba2,rojas2021impact,yang2021probing,Wang21,Banados21,liu2021constraining}. Indeed, the fraction of observed high-$z$ radio loud quasars is lower than theoretically expected (see e.g. \citealt{volonteri2011}), possibly due to a suppression of their radio flux due to CMB quenching (see e.g. \citealt{ghisellini2014radio,ghisellini2015cmb,wu2017cmb}).  While greatly debated, an increasing number of observations and next generation of telescopes promise to answer several of these questions.

\section*{Acknowledgements}
This work is partly funded by Vici grant 639.043.409 from the Dutch Research Council (NWO). The authors would like to thank Nastasha Wijers for her support with the use of the Aurora simulation and constructive comments, Enrico Garaldi, Tiago Costa and Abhijeet Anand for useful discussions,and an anonymous referee for their constructive comments. This work extensively made use of multiple publicly available software packages :  {\tt matplotlib} \citep{hunter2007matplotlib}, {\tt numpy} \cite{van2011numpy}, {\tt scipy} \citep{scipy} and {\tt yt} \citep{YT2011}. The authors would like to thank the community of developers and those maintaining these packages.

\section*{Data availability}
Absorber catalogs and relevant absorber statistics will be shared on reasonable request to the corresponding author.



\bibliographystyle{mnras}
\interlinepenalty=10000
\bibliography{biblio.bib}

\begin{thebibliography}{}
\makeatletter
\relax
\def\mn@urlcharsother{\let\do\@makeother \do\$\do\&\do\#\do\^\do\_\do\%\do\~}
\def\mn@doi{\begingroup\mn@urlcharsother \@ifnextchar [ {\mn@doi@}
  {\mn@doi@[]}}
\def\mn@doi@[#1]#2{\def\@tempa{#1}\ifx\@tempa\@empty \href
  {http://dx.doi.org/#2} {doi:#2}\else \href {http://dx.doi.org/#2} {#1}\fi
  \endgroup}
\def\mn@eprint#1#2{\mn@eprint@#1:#2::\@nil}
\def\mn@eprint@arXiv#1{\href {http://arxiv.org/abs/#1} {{\tt arXiv:#1}}}
\def\mn@eprint@dblp#1{\href {http://dblp.uni-trier.de/rec/bibtex/#1.xml}
  {dblp:#1}}
\def\mn@eprint@#1:#2:#3:#4\@nil{\def\@tempa {#1}\def\@tempb {#2}\def\@tempc
  {#3}\ifx \@tempc \@empty \let \@tempc \@tempb \let \@tempb \@tempa \fi \ifx
  \@tempb \@empty \def\@tempb {arXiv}\fi \@ifundefined
  {mn@eprint@\@tempb}{\@tempb:\@tempc}{\expandafter \expandafter \csname
  mn@eprint@\@tempb\endcsname \expandafter{\@tempc}}}

\bibitem[\protect\citeauthoryear{Asplund, Grevesse  \& Sauval}{Asplund
  et~al.}{2004}]{asplund2004solar}
Asplund M.,  Grevesse N.,   Sauval J.,  2004, arXiv preprint astro-ph/0410214

\bibitem[\protect\citeauthoryear{{Ba{\~n}ados} et~al.,}{{Ba{\~n}ados}
  et~al.}{2021}]{Banados21}
{Ba{\~n}ados} E.,  et~al., 2021, \mn@doi [\apj] {10.3847/1538-4357/abe239},
  \href {https://ui.adsabs.harvard.edu/abs/2021ApJ...909...80B} {909, 80}

\bibitem[\protect\citeauthoryear{Barkana \& Loeb}{Barkana \&
  Loeb}{2001}]{barkana2001}
Barkana R.,  Loeb A.,  2001, Physics reports, 349, 125

\bibitem[\protect\citeauthoryear{Becker, Sargent, Rauch  \& Simcoe}{Becker
  et~al.}{2006}]{becker2006OI}
Becker G.~D.,  Sargent W.~L.,  Rauch M.,   Simcoe R.~A.,  2006, The
  Astrophysical Journal, 640, 69

\bibitem[\protect\citeauthoryear{Becker, Sargent, Rauch  \& Calverley}{Becker
  et~al.}{2011a}]{becker2011highZmetalsII}
Becker G.~D.,  Sargent W.~L.,  Rauch M.,   Calverley A.~P.,  2011a, The
  Astrophysical Journal, 735, 93

\bibitem[\protect\citeauthoryear{Becker, Sargent, Rauch  \& Carswell}{Becker
  et~al.}{2011b}]{becker2011iron}
Becker G.~D.,  Sargent W.~L.,  Rauch M.,   Carswell R.~F.,  2011b, The
  Astrophysical Journal, 744, 91

\bibitem[\protect\citeauthoryear{{Becker}, {Bolton}  \& {Lidz}}{{Becker}
  et~al.}{2015}]{beckerHighzReview}
{Becker} G.~D.,  {Bolton} J.~S.,   {Lidz} A.,  2015, \mn@doi [\pasa]
  {10.1017/pasa.2015.45}, \href
  {https://ui.adsabs.harvard.edu/abs/2015PASA...32...45B} {32, e045}

\bibitem[\protect\citeauthoryear{Becker et~al.,}{Becker
  et~al.}{2019}]{becker2019evolution}
Becker G.~D.,  et~al., 2019, The Astrophysical Journal, 883, 163

\bibitem[\protect\citeauthoryear{Bosman, Becker, Haehnelt, Hewett, McMahon,
  Mortlock, Simpson  \& Venemans}{Bosman et~al.}{2017}]{bosman2017deep}
Bosman S.~E.,  Becker G.~D.,  Haehnelt M.~G.,  Hewett P.~C.,  McMahon R.~G.,
  Mortlock D.~J.,  Simpson C.,   Venemans B.~P.,  2017, Monthly Notices of the
  Royal Astronomical Society, 470, 1919

\bibitem[\protect\citeauthoryear{Braun, Bonaldi, Bourke, Keane  \& Wagg}{Braun
  et~al.}{2019}]{braun2019anticipated}
Braun R.,  Bonaldi A.,  Bourke T.,  Keane E.,   Wagg J.,  2019, arXiv preprint
  arXiv:1912.12699

\bibitem[\protect\citeauthoryear{Carilli}{Carilli}{2006}]{carilli2006hi}
Carilli C.~L.,  2006, New Astronomy Reviews, 50, 162

\bibitem[\protect\citeauthoryear{Carilli, Gnedin  \& Owen}{Carilli
  et~al.}{2002}]{carilli2002hi}
Carilli C.,  Gnedin N.,   Owen F.,  2002, The Astrophysical Journal, 577, 22

\bibitem[\protect\citeauthoryear{Carilli, Gnedin, Furlanetto  \& Owen}{Carilli
  et~al.}{2004}]{carilli2004SKA21cm}
Carilli C.~L.,  Gnedin N.,  Furlanetto S.,   Owen F.,  2004, New Astronomy
  Reviews, 48, 1053

\bibitem[\protect\citeauthoryear{Castro-Tirado et~al.,}{Castro-Tirado
  et~al.}{2013}]{castro2013grb}
Castro-Tirado A.,  et~al., 2013, arXiv preprint arXiv:1312.5631

\bibitem[\protect\citeauthoryear{{Chabrier}}{{Chabrier}}{2003}]{chabrier_2003}
{Chabrier} G.,  2003, \mn@doi [\pasp] {10.1086/376392}, \href
  {http://adsabs.harvard.edu/abs/2003PASP..115..763C} {115, 763}

\bibitem[\protect\citeauthoryear{Chen et~al.,}{Chen et~al.}{2017}]{chen2017mg}
Chen S.-F.~S.,  et~al., 2017, The Astrophysical Journal, 850, 188

\bibitem[\protect\citeauthoryear{Chornock, Berger, Fox, Lunnan, Drout, Fong,
  Laskar  \& Roth}{Chornock et~al.}{2013}]{chornock2013grb}
Chornock R.,  Berger E.,  Fox D.~B.,  Lunnan R.,  Drout M.~R.,  Fong W.-f.,
  Laskar T.,   Roth K.~C.,  2013, The Astrophysical Journal, 774, 26

\bibitem[\protect\citeauthoryear{Chornock, Berger, Fox, Fong, Laskar  \&
  Roth}{Chornock et~al.}{2014}]{chornock2014grb}
Chornock R.,  Berger E.,  Fox D.~B.,  Fong W.-f.,  Laskar T.,   Roth K.~C.,
  2014, arXiv preprint arXiv:1405.7400

\bibitem[\protect\citeauthoryear{Choudhury \& Ferrara}{Choudhury \&
  Ferrara}{2005}]{choudhury2005}
Choudhury T.~R.,  Ferrara A.,  2005, Monthly Notices of the Royal Astronomical
  Society, 361, 577

\bibitem[\protect\citeauthoryear{{Ciardi} \& {Ferrara}}{{Ciardi} \&
  {Ferrara}}{2005}]{ciardiferrara2005}
{Ciardi} B.,  {Ferrara} A.,  2005, \mn@doi [\ssr] {10.1007/s11214-005-3592-0},
  \href {https://ui.adsabs.harvard.edu/abs/2005SSRv..116..625C} {116, 625}

\bibitem[\protect\citeauthoryear{{Ciardi} \& {Madau}}{{Ciardi} \&
  {Madau}}{2003}]{ciardimadau2003}
{Ciardi} B.,  {Madau} P.,  2003, \mn@doi [\apj] {10.1086/377634}, \href
  {https://ui.adsabs.harvard.edu/abs/2003ApJ...596....1C} {596, 1}

\bibitem[\protect\citeauthoryear{{Ciardi} et~al.,}{{Ciardi}
  et~al.}{2013}]{ciardi2013}
{Ciardi} B.,  et~al., 2013, \mn@doi [\mnras] {10.1093/mnras/sts156}, \href
  {https://ui.adsabs.harvard.edu/abs/2013MNRAS.428.1755C} {428, 1755}

\bibitem[\protect\citeauthoryear{{Ciardi} et~al.,}{{Ciardi}
  et~al.}{2015}]{ciardi2015}
{Ciardi} B.,  et~al., 2015, \mn@doi [\mnras] {10.1093/mnras/stv1640}, \href
  {https://ui.adsabs.harvard.edu/abs/2015MNRAS.453..101C} {453, 101}

\bibitem[\protect\citeauthoryear{{Cooper}, {Simcoe}, {Cooksey}, {Bordoloi},
  {Miller}, {Furesz}, {Turner}  \& {Ba{\~n}ados}}{{Cooper}
  et~al.}{2019b}]{Cooper19}
{Cooper} T.~J.,  {Simcoe} R.~A.,  {Cooksey} K.~L.,  {Bordoloi} R.,  {Miller}
  D.~R.,  {Furesz} G.,  {Turner} M.~L.,   {Ba{\~n}ados} E.,  2019b, \mn@doi
  [\apj] {10.3847/1538-4357/ab3402}, \href
  {https://ui.adsabs.harvard.edu/abs/2019ApJ...882...77C} {882, 77}

\bibitem[\protect\citeauthoryear{Cooper, Simcoe, Cooksey, Bordoloi, Miller,
  Furesz, Turner  \& Ba{\~n}ados}{Cooper et~al.}{2019a}]{cooper2019heavy}
Cooper T.~J.,  Simcoe R.~A.,  Cooksey K.~L.,  Bordoloi R.,  Miller D.~R.,
  Furesz G.,  Turner M.~L.,   Ba{\~n}ados E.,  2019a, The Astrophysical
  Journal, 882, 77

\bibitem[\protect\citeauthoryear{D'Odorico et~al.,}{D'Odorico
  et~al.}{2013}]{dodorico2013}
D'Odorico V.,  et~al., 2013, Monthly Notices of the Royal Astronomical Society,
  435, 1198

\bibitem[\protect\citeauthoryear{D'Odorico et~al.,}{D'Odorico
  et~al.}{2022}]{dodoricoSiIV22}
D'Odorico V.,  et~al., 2022, \mn@doi [Monthly Notices of the Royal Astronomical
  Society] {10.1093/mnras/stac545}, 512, 2389

\bibitem[\protect\citeauthoryear{{Dalla Vecchia} \& {Schaye}}{{Dalla Vecchia}
  \& {Schaye}}{2012}]{vecchia_schaye_2012}
{Dalla Vecchia} C.,  {Schaye} J.,  2012, \mn@doi [\mnras]
  {10.1111/j.1365-2966.2012.21704.x}, \href
  {http://adsabs.harvard.edu/abs/2012MNRAS.426..140D} {426, 140}

\bibitem[\protect\citeauthoryear{{Dayal} \& {Ferrara}}{{Dayal} \&
  {Ferrara}}{2018}]{Dayal2018}
{Dayal} P.,  {Ferrara} A.,  2018, \mn@doi [\physrep]
  {10.1016/j.physrep.2018.10.002}, \href
  {https://ui.adsabs.harvard.edu/abs/2018PhR...780....1D} {780, 1}

\bibitem[\protect\citeauthoryear{Doughty \& Finlator}{Doughty \&
  Finlator}{2019a}]{doughty2019evoluOI}
Doughty C.,  Finlator K.,  2019a, Monthly Notices of the Royal Astronomical
  Society

\bibitem[\protect\citeauthoryear{Doughty \& Finlator}{Doughty \&
  Finlator}{2019b}]{doughty2019evolution}
Doughty C.,  Finlator K.,  2019b, Monthly Notices of the Royal Astronomical
  Society, 489, 2755

\bibitem[\protect\citeauthoryear{Doughty, Finlator, Oppenheimer, Dav{\'e}  \&
  Zackrisson}{Doughty et~al.}{2018}]{doughty2018alignedUVB}
Doughty C.,  Finlator K.,  Oppenheimer B.~D.,  Dav{\'e} R.,   Zackrisson E.,
  2018, Monthly Notices of the Royal Astronomical Society, 475, 4717

\bibitem[\protect\citeauthoryear{{Euclid Collaboration} et~al.,}{{Euclid
  Collaboration} et~al.}{2019}]{Barnett19}
{Euclid Collaboration} et~al., 2019, \mn@doi [\aap]
  {10.1051/0004-6361/201936427}, \href
  {https://ui.adsabs.harvard.edu/abs/2019A&A...631A..85E} {631, A85}

\bibitem[\protect\citeauthoryear{Fan et~al.,}{Fan et~al.}{2006}]{fan2006}
Fan X.,  et~al., 2006, The Astronomical Journal, 132, 117

\bibitem[\protect\citeauthoryear{Ferland et~al.,}{Ferland
  et~al.}{2013}]{cloudy}
Ferland G.,  et~al., 2013, Revista mexicana de astronom{\'\i}a y
  astrof{\'\i}sica, 49, 137

\bibitem[\protect\citeauthoryear{Finlator, Mu{\~n}oz, Oppenheimer, Oh, {\"O}zel
   \& Dav{\'e}}{Finlator et~al.}{2013}]{finlator2013hostOI}
Finlator K.,  Mu{\~n}oz J.~A.,  Oppenheimer B.,  Oh S.~P.,  {\"O}zel F.,
  Dav{\'e} R.,  2013, Monthly Notices of the Royal Astronomical Society, 436,
  1818

\bibitem[\protect\citeauthoryear{Finlator, Oppenheimer, Dav{\'e}, Zackrisson,
  Thompson  \& Huang}{Finlator et~al.}{2016}]{finlator2016softUVB}
Finlator K.,  Oppenheimer B.,  Dav{\'e} R.,  Zackrisson E.,  Thompson R.,
  Huang S.,  2016, Monthly Notices of the Royal Astronomical Society, 459, 2299

\bibitem[\protect\citeauthoryear{Furlanetto}{Furlanetto}{2006}]{furlanetto200621cmglobal}
Furlanetto S.~R.,  2006, Monthly Notices of the Royal Astronomical Society,
  371, 867

\bibitem[\protect\citeauthoryear{Furlanetto \& Loeb}{Furlanetto \&
  Loeb}{2002}]{furlanetto200221minihalo}
Furlanetto S.~R.,  Loeb A.,  2002, The Astrophysical Journal, 579, 1

\bibitem[\protect\citeauthoryear{Furlanetto \& Loeb}{Furlanetto \&
  Loeb}{2003}]{furlanetto2003metalabs}
Furlanetto S.~R.,  Loeb A.,  2003, The Astrophysical Journal, 588, 18

\bibitem[\protect\citeauthoryear{Furlanetto \& Mesinger}{Furlanetto \&
  Mesinger}{2009}]{furlanetto2009UVB}
Furlanetto S.~R.,  Mesinger A.,  2009, Monthly Notices of the Royal
  Astronomical Society, 394, 1667

\bibitem[\protect\citeauthoryear{Furlanetto, Oh  \& Briggs}{Furlanetto
  et~al.}{2006}]{furlanetto21cm2006}
Furlanetto S.~R.,  Oh S.~P.,   Briggs F.~H.,  2006, Physics reports, 433, 181

\bibitem[\protect\citeauthoryear{Garc{\'\i}a, Tescari, Ryan-Weber  \&
  Wyithe}{Garc{\'\i}a et~al.}{2017}]{garcia2017metalline}
Garc{\'\i}a L.,  Tescari E.,  Ryan-Weber E.,   Wyithe J.,  2017, Monthly
  Notices of the Royal Astronomical Society, 470, 2494

\bibitem[\protect\citeauthoryear{Ghisellini, Celotti, Tavecchio, Haardt  \&
  Sbarrato}{Ghisellini et~al.}{2014}]{ghisellini2014radio}
Ghisellini G.,  Celotti A.,  Tavecchio F.,  Haardt F.,   Sbarrato T.,  2014,
  Monthly Notices of the Royal Astronomical Society, 438, 2694

\bibitem[\protect\citeauthoryear{Ghisellini, Haardt, Ciardi, Sbarrato, Gallo,
  Tavecchio  \& Celotti}{Ghisellini et~al.}{2015}]{ghisellini2015cmb}
Ghisellini G.,  Haardt F.,  Ciardi B.,  Sbarrato T.,  Gallo E.,  Tavecchio F.,
   Celotti A.,  2015, Monthly Notices of the Royal Astronomical Society, 452,
  3457

\bibitem[\protect\citeauthoryear{Gnedin}{Gnedin}{2000}]{gnedin2000}
Gnedin N.~Y.,  2000, The Astrophysical Journal, 542, 535

\bibitem[\protect\citeauthoryear{Haardt \& Madau}{Haardt \& Madau}{2012}]{HM12}
Haardt F.,  Madau P.,  2012, The Astrophysical Journal, 746, 125

\bibitem[\protect\citeauthoryear{Hartoog et~al.,}{Hartoog
  et~al.}{2015}]{hartoog2015vlt}
Hartoog O.,  et~al., 2015, Astronomy \& Astrophysics, 580, A139

\bibitem[\protect\citeauthoryear{Hennawi, Davies, Wang  \& O{\~n}orbe}{Hennawi
  et~al.}{2021}]{hennawi2021probing}
Hennawi J.~F.,  Davies F.~B.,  Wang F.,   O{\~n}orbe J.,  2021, Monthly Notices
  of the Royal Astronomical Society, 506, 2963

\bibitem[\protect\citeauthoryear{{Hern{\'a}ndez-Monteagudo}, {Haiman},
  {Jimenez}  \& {Verde}}{{Hern{\'a}ndez-Monteagudo} et~al.}{2007}]{Oxypump2007}
{Hern{\'a}ndez-Monteagudo} C.,  {Haiman} Z.,  {Jimenez} R.,   {Verde} L.,
  2007, \mn@doi [\apjl] {10.1086/518090}, \href
  {http://adsabs.harvard.edu/abs/2007ApJ...660L..85H} {660, L85}

\bibitem[\protect\citeauthoryear{{Hummels}, {Smith}  \& {Silvia}}{{Hummels}
  et~al.}{2017}]{Trident2017}
{Hummels} C.~B.,  {Smith} B.~D.,   {Silvia} D.~W.,  2017, \mn@doi [\apj]
  {10.3847/1538-4357/aa7e2d}, \href
  {http://adsabs.harvard.edu/abs/2017ApJ...847...59H} {847, 59}

\bibitem[\protect\citeauthoryear{Hunter}{Hunter}{2007}]{hunter2007matplotlib}
Hunter J.~D.,  2007, Computing in science \& engineering, 9, 90

\bibitem[\protect\citeauthoryear{Jones, Oliphant, Peterson  et~al.}{Jones
  et~al.}{01  }]{scipy}
Jones E.,  Oliphant T.,  Peterson P.,   et~al., 2001--, {SciPy}: Open source
  scientific tools for {Python}, \url {http://www.scipy.org/}

\bibitem[\protect\citeauthoryear{Keating, Haehnelt, Becker  \& Bolton}{Keating
  et~al.}{2013}]{keating2013OI}
Keating L.~C.,  Haehnelt M.~G.,  Becker G.~D.,   Bolton J.~S.,  2013, Monthly
  Notices of the Royal Astronomical Society, 438, 1820

\bibitem[\protect\citeauthoryear{{Komatsu} et~al.,}{{Komatsu}
  et~al.}{2011}]{komatsu_2011}
{Komatsu} E.,  et~al., 2011, \mn@doi [\apjs] {10.1088/0067-0049/192/2/18},
  \href {http://adsabs.harvard.edu/abs/2011ApJS..192...18K} {192, 18}

\bibitem[\protect\citeauthoryear{Li et~al.,}{Li et~al.}{2020}]{li2020scuba2}
Li Q.,  et~al., 2020, The Astrophysical Journal, 900, 12

\bibitem[\protect\citeauthoryear{Liu et~al.,}{Liu
  et~al.}{2021}]{liu2021constraining}
Liu Y.,  et~al., 2021, The Astrophysical Journal, 908, 124

\bibitem[\protect\citeauthoryear{Loeb \& Furlanetto}{Loeb \&
  Furlanetto}{2013}]{firstgalaxiesloeb13}
Loeb A.,  Furlanetto S.~R.,  2013, The first galaxies in the universe.
Princeton University Press

\bibitem[\protect\citeauthoryear{Madau, Meiksin  \& Rees}{Madau
  et~al.}{1997}]{madau199721cm}
Madau P.,  Meiksin A.,   Rees M.~J.,  1997, The Astrophysical Journal, 475, 429

\bibitem[\protect\citeauthoryear{{Marconi} et~al.,}{{Marconi}
  et~al.}{2021}]{Marconi21}
{Marconi} A.,  et~al., 2021, \mn@doi [The Messenger] {10.18727/0722-6691/5219},
  \href {https://ui.adsabs.harvard.edu/abs/2021Msngr.182...27M} {182, 27}

\bibitem[\protect\citeauthoryear{McGreer, Mesinger  \& D'Odorico}{McGreer
  et~al.}{2014}]{mcgreer2014reion}
McGreer I.~D.,  Mesinger A.,   D'Odorico V.,  2014, Monthly Notices of the
  Royal Astronomical Society, 447, 499

\bibitem[\protect\citeauthoryear{Meiksin}{Meiksin}{2009}]{meiksin2009IGM}
Meiksin A.~A.,  2009, Reviews of modern physics, 81, 1405

\bibitem[\protect\citeauthoryear{Mesinger}{Mesinger}{2010}]{mesinger2010reion}
Mesinger A.,  2010, Monthly Notices of the Royal Astronomical Society, 407,
  1328

\bibitem[\protect\citeauthoryear{Meyer, Bosman  \& Ellis}{Meyer
  et~al.}{2019}]{meyer2019new}
Meyer R.~A.,  Bosman S.~E.,   Ellis R.~S.,  2019, Monthly Notices of the Royal
  Astronomical Society, 487, 3305

\bibitem[\protect\citeauthoryear{Mort, Dulwich, Salvini, Adami  \& Jones}{Mort
  et~al.}{2010}]{mort2010oskar}
Mort B.~J.,  Dulwich F.,  Salvini S.,  Adami K.~Z.,   Jones M.~E.,  2010, in
  2010 IEEE International Symposium on Phased Array Systems and Technology. pp
  690--694

\bibitem[\protect\citeauthoryear{Mortlock et~al.,}{Mortlock
  et~al.}{2011}]{mortlock2011quasar}
Mortlock D.~J.,  et~al., 2011, Nature, 474, 616

\bibitem[\protect\citeauthoryear{{Oh}}{{Oh}}{2002}]{oh2002}
{Oh} S.~P.,  2002, \mn@doi [\mnras] {10.1046/j.1365-8711.2002.05859.x}, \href
  {https://ui.adsabs.harvard.edu/\#abs/2002MNRAS.336.1021O} {336, 1021}

\bibitem[\protect\citeauthoryear{Oppenheimer, Dav{\'e}  \&
  Finlator}{Oppenheimer et~al.}{2009}]{BenOpp2009metallines}
Oppenheimer B.~D.,  Dav{\'e} R.,   Finlator K.,  2009, Monthly Notices of the
  Royal Astronomical Society, 396, 729

\bibitem[\protect\citeauthoryear{{Osterbrock}}{{Osterbrock}}{1989}]{Osterbrock}
{Osterbrock} D.~E.,  1989, {Astrophysics of gaseous nebulae and active galactic
  nuclei}

\bibitem[\protect\citeauthoryear{{Pawlik} \& {Schaye}}{{Pawlik} \&
  {Schaye}}{2008}]{pawlik_schaye_2008}
{Pawlik} A.~H.,  {Schaye} J.,  2008, \mn@doi [\mnras]
  {10.1111/j.1365-2966.2008.13601.x}, \href
  {http://adsabs.harvard.edu/abs/2008MNRAS.389..651P} {389, 651}

\bibitem[\protect\citeauthoryear{{Pawlik} \& {Schaye}}{{Pawlik} \&
  {Schaye}}{2011}]{pawlik_schaye_2011}
{Pawlik} A.~H.,  {Schaye} J.,  2011, \mn@doi [\mnras]
  {10.1111/j.1365-2966.2010.18032.x}, \href
  {http://adsabs.harvard.edu/abs/2011MNRAS.412.1943P} {412, 1943}

\bibitem[\protect\citeauthoryear{Pawlik, Milosavljevi{\'c}  \& Bromm}{Pawlik
  et~al.}{2013}]{pawlik_2013}
Pawlik A.~H.,  Milosavljevi{\'c} M.,   Bromm V.,  2013, The Astrophysical
  Journal, 767, 59

\bibitem[\protect\citeauthoryear{{Pawlik}, {Rahmati}, {Schaye}, {Jeon}  \&
  {Dalla Vecchia}}{{Pawlik} et~al.}{2017}]{Pawlik_etal_2017}
{Pawlik} A.~H.,  {Rahmati} A.,  {Schaye} J.,  {Jeon} M.,   {Dalla Vecchia} C.,
  2017, \mn@doi [\mnras] {10.1093/mnras/stw2869}, \href
  {http://adsabs.harvard.edu/abs/2017MNRAS.466..960P} {466, 960}

\bibitem[\protect\citeauthoryear{Pritchard \& Loeb}{Pritchard \&
  Loeb}{2008}]{pritchard200821cm}
Pritchard J.~R.,  Loeb A.,  2008, Physical Review D, 78, 103511

\bibitem[\protect\citeauthoryear{{Rai{\v c}evi{\'c}}, {Pawlik}, {Schaye}  \&
  {Rahmati}}{{Rai{\v c}evi{\'c}} et~al.}{2014}]{raicevic_2014}
{Rai{\v c}evi{\'c}} M.,  {Pawlik} A.~H.,  {Schaye} J.,   {Rahmati} A.,  2014,
  \mn@doi [\mnras] {10.1093/mnras/stt2099}, \href
  {http://adsabs.harvard.edu/abs/2014MNRAS.437.2816R} {437, 2816}

\bibitem[\protect\citeauthoryear{Reed et~al.,}{Reed
  et~al.}{2015}]{reed2015j0454}
Reed S.~L.,  et~al., 2015, Monthly Notices of the Royal Astronomical Society,
  454, 3952

\bibitem[\protect\citeauthoryear{Rojas-Ruiz et~al.,}{Rojas-Ruiz
  et~al.}{2021}]{rojas2021impact}
Rojas-Ruiz S.,  et~al., 2021, arXiv preprint arXiv:2108.04257

\bibitem[\protect\citeauthoryear{Ryan-Weber, Pettini  \& Madau}{Ryan-Weber
  et~al.}{2006}]{WeberRyan2006CIVabs}
Ryan-Weber E.~V.,  Pettini M.,   Madau P.,  2006, Monthly Notices of the Royal
  Astronomical Society: Letters, 371, L78

\bibitem[\protect\citeauthoryear{Ryan-Weber, Pettini, Madau  \&
  Zych}{Ryan-Weber et~al.}{2009}]{WeberRyan2009downturnCIV}
Ryan-Weber E.~V.,  Pettini M.,  Madau P.,   Zych B.~J.,  2009, Monthly Notices
  of the Royal Astronomical Society, 395, 1476

\bibitem[\protect\citeauthoryear{{Schaerer}}{{Schaerer}}{2003}]{schaerer_2003}
{Schaerer} D.,  2003, \mn@doi [\aap] {10.1051/0004-6361:20021525}, \href
  {http://adsabs.harvard.edu/abs/2003A%26A...397..527S} {397, 527}

\bibitem[\protect\citeauthoryear{{Schaye} \& {Dalla Vecchia}}{{Schaye} \&
  {Dalla Vecchia}}{2008}]{schaye_vecchia_2008}
{Schaye} J.,  {Dalla Vecchia} C.,  2008, \mn@doi [\mnras]
  {10.1111/j.1365-2966.2007.12639.x}, \href
  {http://adsabs.harvard.edu/abs/2008MNRAS.383.1210S} {383, 1210}

\bibitem[\protect\citeauthoryear{{Semelin}}{{Semelin}}{2016}]{semelin_2016}
{Semelin} B.,  2016, \mn@doi [\mnras] {10.1093/mnras/stv2312}, \href
  {http://adsabs.harvard.edu/abs/2016MNRAS.455..962S} {455, 962}

\bibitem[\protect\citeauthoryear{Simcoe}{Simcoe}{2006}]{simcoe2006CIV}
Simcoe R.~A.,  2006, The Astrophysical Journal, 653, 977

\bibitem[\protect\citeauthoryear{Simcoe et~al.,}{Simcoe
  et~al.}{2011}]{simcoe2011CIVdensity}
Simcoe R.~A.,  et~al., 2011, The Astrophysical Journal, 743, 21

\bibitem[\protect\citeauthoryear{{\v{S}}oltinsk{\`y}
  et~al.,}{{\v{S}}oltinsk{\`y} et~al.}{2021}]{vsoltinsky2021}
{\v{S}}oltinsk{\`y} T.,  et~al., 2021, arXiv preprint arXiv:2105.02250

\bibitem[\protect\citeauthoryear{{Springel}}{{Springel}}{2005}]{Springel_2005}
{Springel} V.,  2005, \mn@doi [\mnras] {10.1111/j.1365-2966.2005.09655.x},
  \href {http://adsabs.harvard.edu/abs/2005MNRAS.364.1105S} {364, 1105}

\bibitem[\protect\citeauthoryear{{Turk}, {Smith}, {Oishi}, {Skory}, {Skillman},
  {Abel}  \& {Norman}}{{Turk} et~al.}{2011}]{YT2011}
{Turk} M.~J.,  {Smith} B.~D.,  {Oishi} J.~S.,  {Skory} S.,  {Skillman} S.~W.,
  {Abel} T.,   {Norman} M.~L.,  2011, \mn@doi [The Astrophysical Journal
  Supplement Series] {10.1088/0067-0049/192/1/9}, \href
  {https://ui.adsabs.harvard.edu/abs/2011ApJS..192....9T} {192, 9}

\bibitem[\protect\citeauthoryear{Van Der~Walt, Colbert  \& Varoquaux}{Van
  Der~Walt et~al.}{2011}]{van2011numpy}
Van Der~Walt S.,  Colbert S.~C.,   Varoquaux G.,  2011, Computing in science \&
  engineering, 13, 22

\bibitem[\protect\citeauthoryear{{Volonteri}, {Haardt}, {Ghisellini}  \& {Della
  Ceca}}{{Volonteri} et~al.}{2011}]{volonteri2011}
{Volonteri} M.,  {Haardt} F.,  {Ghisellini} G.,   {Della Ceca} R.,  2011,
  \mn@doi [\mnras] {10.1111/j.1365-2966.2011.19024.x}, \href
  {https://ui.adsabs.harvard.edu/abs/2011MNRAS.416..216V} {416, 216}

\bibitem[\protect\citeauthoryear{{Wang} et~al.,}{{Wang} et~al.}{2021}]{Wang21}
{Wang} F.,  et~al., 2021, \mn@doi [\apjl] {10.3847/2041-8213/abd8c6}, \href
  {https://ui.adsabs.harvard.edu/abs/2021ApJ...907L...1W} {907, L1}

\bibitem[\protect\citeauthoryear{Wiersma, Schaye, Theuns, Dalla~Vecchia  \&
  Tornatore}{Wiersma et~al.}{2009}]{wiersma2009}
Wiersma R.~P.,  Schaye J.,  Theuns T.,  Dalla~Vecchia C.,   Tornatore L.,
  2009, Monthly Notices of the Royal Astronomical Society, 399, 574

\bibitem[\protect\citeauthoryear{Wu, Ghisellini, Hodges-Kluck, Gallo, Ciardi,
  Haardt, Sbarrato  \& Tavecchio}{Wu et~al.}{2017}]{wu2017cmb}
Wu J.,  Ghisellini G.,  Hodges-Kluck E.,  Gallo E.,  Ciardi B.,  Haardt F.,
  Sbarrato T.,   Tavecchio F.,  2017, Monthly Notices of the Royal Astronomical
  Society, 468, 109

\bibitem[\protect\citeauthoryear{Wu et~al.,}{Wu et~al.}{2021}]{wu2021oi}
Wu Y.,  et~al., 2021, Nature Astronomy, pp~1--8

\bibitem[\protect\citeauthoryear{Xu, Chen, Fan, Trac  \& Cen}{Xu
  et~al.}{2009}]{xu200921cm}
Xu Y.,  Chen X.,  Fan Z.,  Trac H.,   Cen R.,  2009, The Astrophysical Journal,
  704, 1396

\bibitem[\protect\citeauthoryear{Xu, Ferrara  \& Chen}{Xu
  et~al.}{2011}]{xu201121cmgal}
Xu Y.,  Ferrara A.,   Chen X.,  2011, Monthly Notices of the Royal Astronomical
  Society, 410, 2025

\bibitem[\protect\citeauthoryear{Yang et~al.,}{Yang
  et~al.}{2021}]{yang2021probing}
Yang J.,  et~al., 2021, Probing Early Super-massive Black Hole Growth and
  Quasar Evolution with Near-infrared Spectroscopy of 37 Reionization-era
  Quasars at 6.3 < z <= 7.64 (\mn@eprint {arXiv} {2109.13942})

\bibitem[\protect\citeauthoryear{Zeimann, White, Becker, Hodge, Stanford  \&
  Richards}{Zeimann et~al.}{2011}]{zeimann2011discovery}
Zeimann G.~R.,  White R.~L.,  Becker R.~H.,  Hodge J.~A.,  Stanford S.~A.,
  Richards G.~T.,  2011, The Astrophysical Journal, 736, 57

\makeatother
\end{thebibliography}





\bsp	
\label{lastpage}
\end{document}